\documentclass[aps,prx,twocolumn,showpacs]{revtex4-1}
\usepackage{bm}
\usepackage{mathrsfs}
\usepackage{amsmath}
\usepackage{amssymb}
\usepackage{graphicx}
\usepackage{amsfonts}
\usepackage{amsthm}
\usepackage{color}
\usepackage{dcolumn}
\usepackage{txfonts}
\usepackage[caption=false]{subfig}
\usepackage[T1]{fontenc}
\usepackage{multirow}

\DeclareMathOperator{\sn}{sn}
\DeclareMathOperator{\cn}{cn}
\DeclareMathOperator{\dn}{dn}
\DeclareMathOperator{\sd}{sd}
\DeclareMathOperator{\cd}{cd}
\DeclareMathOperator{\nd}{nd}

\DeclareMathOperator{\sech}{sech}

\begin{document}

\title{Nonlinear optical analogues of quantum phase transitions in a squeezing-enhanced LMG model}
\author{Chon-Fai Kam}
\email{dubussygauss@gmail.com}
\affiliation{Department of Physics, University at Buffalo, SUNY, Buffalo, New York 14260-1500, USA}

\begin{abstract}
We investigate nonlinear optical analogues of quantum phase transitions within a squeezing-enhanced generalized Lipkin-Meshkov-Glick (LMG) model, focusing on excited-state quantum phase transitions in optical fibers with tetragonal symmetry. Our analysis reveals a novel squeezing effect that induces classical bifurcations in polarization dynamics, even without a linear rotor-like term. By mapping the nonlinear polarization dynamics to the generalized LMG model, we establish a direct correspondence between optical bifurcations and quantum critical phenomena, uncovering geometric gauge structures akin to Berry-like phases. These findings highlight the interplay between classical and quantum behaviors in optical systems, offering a versatile platform for studying quantum many-body physics with applications in quantum metrology and simulation.
\end{abstract}

\maketitle

\section{Introduction}\label{1}
The Lipkin-Meshkov-Glick (LMG) model \cite{lipkin1965validity}, one of the earliest and most extensively studied models for quantum phase transitions, was initially developed for nuclear systems and later applied to Bose-Einstein condensates in cold atomic systems \cite{milburn1997quantum}. Alongside it, the quantum rotor model, used to describe systems like superfluids and magnets, is another key framework \cite{sachdev2011quantum}. Both models exhibit complex phase diagrams and universal scaling behavior near critical points \cite{cejnar2010quantum}. Recently, theoretical studies have explored the generalized LMG model, which effectively integrates features of both the LMG and quantum rotor models \cite{opatrny2018analogies}. In particular, the generalized LMG model extends the standard LMG framework by incorporating rotor-like degrees of freedom, allowing it to capture a broader range of quantum critical phenomena, including spin squeezing \cite{ma2011quantum}, excited-state quantum phase transitions (QPTs) \cite{caprio2008excited}, and time-crystalline behavior \cite{russomanno2017floquet}. This synthesis draws inspiration from classical analogs, such as the Euler top with a rotor, where the interplay of collective spin dynamics from the LMG model and rotational degrees of freedom from the quantum rotor model mirrors complex classical motions like precession, nutation, or chaotic tumbling \cite{opatrny2018analogies}. 

The generalized LMG model’s Hamiltonian includes terms that couple the collective spin to additional modes \cite{ferreira2019lipkin}, enabling it to describe systems ranging from interacting spins in quantum simulators to bosonic systems in optical lattices \cite{milburn1997quantum, bloch2008many}. Its phase diagram is enriched by the competition between long-range interactions, external fields, and rotational dynamics \cite{cejnar2010quantum}, leading to multiple critical points and novel phases, such as those analogous to self-trapping in Bose-Einstein condensates \cite{raghavan1999coherent} or Floquet time crystals under periodic driving \cite{russomanno2017floquet}. The universal scaling behavior near these critical points, characterized by mean-field exponents due to the infinite-range interactions, is complemented by finite-size scaling effects \cite{dusuel2004finite, dusuel2005continuous} that are experimentally accessible in cold atomic platforms \cite{bloch2008many}. Recent studies have also leveraged classical-quantum analogies to propose new types of excited-state quantum phase transitions, where singularities in the energy spectrum correspond to bifurcations in the classical Euler top’s phase space \cite{cejnar2010quantum, opatrny2018analogies, kam20172+}.

The study of quantum phase transitions and their associated gauge field singularities has traditionally been confined to purely quantum mechanical systems, such as condensed matter or atomic ensembles. However, nonlinear optical systems, particularly those involving optical fibers, offer a remarkably versatile and experimentally accessible platform for emulating these phenomena \cite{agrawal2000nonlinear}. The evolution of the polarization vector in a nonlinear optical fiber provides a direct analogy to unitary Hamiltonian quantum dynamics, where the optical axis of the fiber serves as the temporal coordinate, replacing the time variable in traditional atomic systems \cite{kam2025nonlinear}.

In such systems, the polarization state of light evolves according to a nonlinear Schrödinger equation, which governs the interaction between the electric field components and the nonlinear susceptibility of the fiber medium \cite{agrawal2000nonlinear}. This evolution can exhibit bifurcations in the polarization dynamics, mirroring the critical points observed in excited-state QPTs \cite{cejnar2010quantum}. These bifurcations arise due to the interplay between linear birefringence, nonlinear Kerr effects, and dispersion, leading to abrupt changes in the polarization state as a function of propagation distance or input power. Remarkably, these optical bifurcations are not merely phenomenological but exhibit a one-to-one correspondence with their quantum counterparts, including the emergence of geometric gauge structures \cite{kam20172+, kam2025nonlinear}.

The geometric gauge structure in these systems manifests as a Berry-like phase \cite{berry1984quantal} or a non-Abelian gauge field \cite{wilczek1989geometric}, which accompanies the adiabatic evolution of the polarization vector along the fiber. This is evident in systems where the optical fiber is engineered to possess spatially varying birefringence or twist, inducing a synthetic gauge field that mimics the topological features of quantum systems \cite{kam2025nonlinear}. Such gauge fields are directly tied to the singularities observed in QPTs, where the energy landscape of the system undergoes a topological reconfiguration \cite{kam20172+}. In the optical context, these singularities appear as critical points in the polarization state space, where the system transitions between distinct dynamical regimes.

Unlike quantum systems, which often require ultra-low temperatures or complex trapping mechanisms, optical fibers operate at room temperature and leverage well-established technologies such as laser sources, modulators, and detectors. By carefully designing the fiber’s properties—such as its core diameter, doping, or periodic structuring—researchers can tailor the effective Hamiltonian governing the polarization dynamics to emulate specific quantum models \cite{russell2003photonic}, including those exhibiting exotic QPTs \cite{li2022high}. For instance, a fiber with a periodically modulated refractive index can simulate a Floquet system, where time-periodic driving in quantum systems is replaced by spatial periodicity along the fiber \cite{ma2018floquet}.

Furthermore, the scalability of optical systems allows for the exploration of higher-dimensional or multipartite quantum analogues. By coupling multiple optical fibers or using multicore fibers \cite{rojas2024non}, one can emulate interacting quantum systems, where the coupling between polarization states in different cores mimics spin-spin interactions or many-body effects. This opens the door to studying collective phenomena, such as optical analogues of quantum entanglement or topological order, within a controlled laboratory setting.

In this work, we aim to systematically investigate the intricate relationship between nonlinear polarization dynamics in nonlinear optical fibers and their correspondence to excited-state QPTs as described by the generalized LMG model. The polarization dynamics along a nonlinear optical fiber exhibit rich behaviors that can be classified according to distinct symmetry classes. These symmetries dictate the structure of the effective Hamiltonian and the nature of the critical points separating different dynamical regimes. By mapping the nonlinear fiber optical system to the generalized LMG model, we reveal deep analogies between classical nonlinear optics and quantum many-body systems, enabling the study of QPT-like phenomena in a highly controllable experimental platform.

\section{Polarization effects in isotropic fibers}\label{2}
To build a solid framework for studying nonlinear optical analogues of quantum phase transitions in interacting quantum systems, we investigate birefringence effects in nonlinear weakly guiding polarization-maintaining fibers in this section. We derive a set of nonlinear Schrödinger equations that characterize the coherent interaction between two orthogonally polarized modes. As in the derivation of Eq.\:\eqref{IsotropicElectricField}, the optical field is assumed to be quasi-monochromatic. This allows us to represent the electric field as the product of a slowly varying function in time and a rapidly oscillating sinusoid. However, unlike isotropic fibers, polarization-maintaining fibers inherently support two non-degenerate polarized modes \cite{agrawal2000nonlinear}. Consequently, the electric field must be expressed in a manner that accounts for these distinct polarization states.
\begin{equation}\label{PolarizedElectricField}
\mathbf{E}(\mathbf{r},t) \equiv \frac{1}{2}\left\{E_x(\mathbf{r},t)\hat{x}+E_y(\mathbf{r},t)\hat{y}\right\}e^{-i\omega_0t}+c.c.,
\end{equation}
where the axial electric field $E_z(\mathbf{r},t)$ is assumed to remain negligible during the entire fiber length. Similarly, we may express the nonlinear part of the induced electric polarization as the product of a slowly varying function in time and a fast oscillating sinusoid by writing
\begin{equation}\label{NewPolarization}
\mathbf{P}_{NL}(\mathbf{r},t) \equiv \frac{1}{2}\left\{P^{NL}_x(\mathbf{r},t)\hat{x}+P^{NL}_y(\mathbf{r},t)\hat{y}\right\}e^{-i\omega_0t}+c.c.
\end{equation}
For sufficiently weak electric fields in media that possess inversion symmetry at the molecular level, such as silica glasses, the induced nonlinear polarization is solely determined by the third-order nonlinear susceptibility. Similar to the reasoning applied to Eq.\:\eqref{ThridOrderElectricPolarization}, the nonlinear response is assumed to be instantaneous, so that the time dependence in the third-order nonlinear susceptibility is evaluated by the product of three delta functions. For such a case, the nonlinear polarization has the form \cite{agrawal2000nonlinear}
\begin{equation}\label{SimplifiedPolarization}
\mathbf{P}_{NL}(\mathbf{r},t)=\epsilon_0\chi^{(3)}\vdots\mathbf{E}(\mathbf{r},t)\mathbf{E}(\mathbf{r},t)\mathbf{E}(\mathbf{r},t),
\end{equation}
where, in general, the third-order susceptibility $\chi^{(3)}$ is a fourth-rank tensor with 81 elements. Substitution of Eqs.\:\eqref{PolarizedElectricField} and \eqref{NewPolarization} into Eq.\:\eqref{SimplifiedPolarization} yields
\begin{align}
&\mathbf{P}_{NL}(\mathbf{r},t)\cdot\hat{i}=\frac{\epsilon_0}{8}\sum_{jkl}\{\chi^{(3)}_{ijkl}E_j(\mathbf{r},t)E_k(\mathbf{r},t)E_l(\mathbf{r},t)e^{-3i\omega_0t}\nonumber\\
+&(\chi^{(3)}_{ijkl}+\chi^{(3)}_{iljk}+\chi^{(3)}_{iklj})E_j(\mathbf{r},t)E_k(\mathbf{r},t)E_l^*(\mathbf{r},t)e^{-i\omega_0t}\}+c.c.
\end{align}
Similar to the reasoning applied to Eq.\:\eqref{ScalarNonlinearPolarization}, the term proportional to $e^{-3i\omega_0t}$ corresponds to frequency tripling processes, which can be neglected if phase-matching techniques are not employed. Hence, the slowly varying function of the nonlinear polarization can be approximately written as
\begin{equation}
P_i^{NL}(\mathbf{r},t)=\frac{\epsilon_0}{4}\sum_{jkl}(\chi^{(3)}_{ijkl}+\chi^{(3)}_{iljk}+\chi^{(3)}_{iklj})E_j(\mathbf{r},t)E_k(\mathbf{r},t)E_l^*(\mathbf{r},t).
\end{equation}
For isotropic media, the third-order nonlinear susceptibility has only three non-zero independent elements, which can be selected as $\chi_{xxyy}^{(3)}$, $\chi_{xyxy}^{(3)}$ and $\chi_{xyyx}^{(3)}$, and the remaining non-zero elements of the third-order nonlinear susceptibility obey the following relationships \cite{boyd2003nonlinear, sutherland2003handbook}
\begin{subequations}
\begin{gather}
\chi_{xxxx}^{(3)}=\chi_{yyyy}^{(3)}=\chi_{xxyy}^{(3)}+\chi_{xyxy}^{(3)}+\chi_{xyyx}^{(3)},\\
\chi_{xxyy}^{(3)}=\chi_{yyxx}^{(3)},\chi_{xyxy}^{(3)}=\chi_{yxyx}^{(3)},\chi_{xyyx}^{(3)}=\chi_{yxxy}^{(3)}.
\end{gather}
\end{subequations}
Hence, the slowly varying function of the nonlinear polarization can be simply written as
\begin{subequations}
\begin{align}\label{IsotropicNonlinearPolarization1}
P_x^{NL}&=\frac{3\epsilon_0}{4}\chi_{xxxx}^{(3)}\left\{\left(|E_x|^2+\frac{2}{3}|E_y|^2\right)E_x+\frac{1}{3}E_y^2E_x^*\right\},\\
P_y^{NL}&=\frac{3\epsilon_0}{4}\chi_{xxxx}^{(3)}\left\{\left(|E_y|^2+\frac{2}{3}|E_x|^2\right)E_x+\frac{1}{3}E_x^2E_y^*\right\}.\label{IsotropicNonlinearPolarization2}
\end{align}
\end{subequations}
In the weakly guiding limit where $\Delta\equiv (n_1-n_c)/n_1\ll 1$, the optical fiber supports a fundamental guided mode, HE$_{11}$, which is resolved into two plane polarized components along the $x$ and $y$ axes in a polarization-maintaining fiber \cite{snyder1983single}. Using the same reasoning applied to Eq.\:\eqref{ModalExpansion}, we may assume that the spatial distribution of the modal field has not been significantly disturbed by nonlinear effects, and hence we may still employ the method of separation of variables, so that the slowly varying functions of the electric field can be expressed in the form
\begin{subequations}
\begin{gather}\label{TwoPolarizedModeAnsatz1}
E_x(\mathbf{r},t)=A_x(z,t)F_x(x,y)e^{i\beta_{0x} z},\\
E_y(\mathbf{r},t)=A_y(z,t)F_y(x,y)e^{i\beta_{0y} z},\label{TwoPolarizedModeAnsatz2}
\end{gather}
\end{subequations}
where $A_x(z,t)$ and $A_y(z,t)$ are the slowly varying modal amplitudes of the two polarized components, $F_x(x,y)$ and $F_y(x,y)$, are the transverse distributions of the fundamental HE$_{11}$ mode of a fiber with elliptical core cross-section or intentional asymmetric stress distribution, and $\beta_{0x}$ and $\beta_{0y}$ are the corresponding propagation constants in the absence of dispersion and nonlinearity. After substituting Eqs.\:\eqref{IsotropicNonlinearPolarization1} and \eqref{IsotropicNonlinearPolarization2} into Maxwells's equations Eq.\:\eqref{MaxwellEquationTime}, the Fourier transform $\tilde{E}_j(\mathbf{r},\omega-\omega_0)$ of $E_j(\mathbf{r},t)$ is found to obey
\begin{subequations}
\begin{gather}\label{MaxwellForTwoPolarizedMode1}
\nabla^2 \tilde{E}_x + \epsilon k^2  \tilde{E}_x +\frac{3k^2}{4}\chi_{xxxx}^{(3)}\left[\widetilde{|E_x|^2E_x}+\frac{2}{3}\widetilde{|E_y|^2E_x}+\frac{1}{3}\widetilde{E_y^2E_x^*}\right]=0,\\
\nabla^2 \tilde{E}_y + \epsilon k^2  \tilde{E}_y +\frac{3k^2}{4}\chi_{xxxx}^{(3)}\left[\widetilde{|E_y|^2E_y}+\frac{2}{3}\widetilde{|E_x|^2E_y}+\frac{1}{3}\widetilde{E_x^2E_y^*}\right]=0,\label{MaxwellForTwoPolarizedMode2}
\end{gather}
\end{subequations}
where $k\equiv \omega/c$ and $\epsilon(\omega)\equiv 1+\tilde{\chi}_{xx}^{(1)}(\omega)$ is the dielectric function. Substituting Eq.\:\eqref{TwoPolarizedModeAnsatz1} into Eq.\:\eqref{MaxwellForTwoPolarizedMode1}, and then multiplying both sides by $F_x^*$ and integrating over $x$ and $y$, we obtain
\begin{align}
&i\frac{\partial \tilde{A}_x}{\partial z}+(\beta_x-\beta_{0x})\tilde{A}_x+i\alpha_x\tilde{A}_x \nonumber\\ 
&+\gamma_x\left(\widetilde{|A_x|^2A_x}+\frac{2}{3}\widetilde{|A_y|^2A_x}+\frac{1}{3}\widetilde{A_y^2A_x^*}e^{-2i\Delta\beta z}\right)=0,
\end{align}
where $\alpha_x\equiv \frac{k_0^2}{2\beta_{0x}}\mbox{Im}\tilde{\chi}_{xx}^{(1)}(\omega_0)$, $\Delta\beta\equiv \beta_{0x}-\beta_{0y}$ is the difference between the propagation constants and $\gamma_x$ is a nonlinearity parameter defined by
\begin{equation}
\gamma_x(\omega_0) \equiv \frac{3k_0}{8n_{e,x}}\chi^{(3)}_{xxxx}\int_{-\infty}^{\infty}\int_{-\infty}^{\infty}|F_x|^4 dxdy\Big/\int_{-\infty}^{\infty}\int_{-\infty}^{\infty}|F_x|^2 dxdy,
\end{equation}
where $n_{e,x}\equiv \beta_{0x}/k_0$ is the effective refractive index of the polarized component along the $x$-direction. Following the same method as in our derivation of Eq.\:\eqref{NonlinearSchrodinger}, we obtain the nonlinear equation that governs the evolution of the complex modal amplitude $A_x(z,t)$
\begin{align}\label{GeneralNLSE1}
&i\frac{\partial A_x}{\partial z}+i\beta_{1x}\frac{\partial A_x}{\partial t}-\frac{\beta_{2x}}{2}\frac{\partial^2 A_x}{\partial t^2}+\frac{i\alpha_x}{2}A_x \nonumber\\ 
&+\gamma_x\left(|A_x|^2A_x+\frac{2}{3}|A_y|^2A_x+\frac{1}{3}A_y^2A_x^*e^{-2i\Delta\beta z}\right)=0,
\end{align}
where $\beta_{1x}$ and $\beta_{2x}$ are the first and second order derivatives of $\beta_x$ evaluated at the carrier frequency $\omega_0$ respectively. The term proportional to $|A_x|^2A_x$ corresponds to the self-phase modulation (SPM) effects \cite{stolen1978self, agrawal2000nonlinear} in an optical fiber, with which the optical field modifies its own phase as a pulse propagates along the fiber; the term proportional to $|A_y|^2A_x$ corresponds to the cross-phase modulation (XPM) effects \cite{agrawal1987modulation}, which results in modulation instability effects \cite{wabnitz1988modulational} --- breakup of continuous-wave (CW) radiation into a train of ultrashort pulses in the case of anomalous group-velocity dispersion ($\beta_{2x}<0$); finally the term proportional to $A_y^2A_x^*$ corresponds to degenerate four-wave mixing (FWM) effects \cite{lin2004vector, agrawal2000nonlinear}. Applying similar procedures to the polarization component along the $y$ axis, we obtain
\begin{align}\label{GeneralNLSE2}
&i\frac{\partial A_y}{\partial z}+i\beta_{1y}\frac{\partial A_y}{\partial t}-\frac{\beta_{2y}}{2}\frac{\partial^2 A_y}{\partial t^2}+\frac{i\alpha_y}{2}A_y \nonumber\\ 
&+\gamma_y\left(|A_y|^2A_y+\frac{2}{3}|A_x|^2A_y+\frac{1}{3}A_x^2A_y^*e^{2i\Delta\beta z}\right)=0.
\end{align}
The terms proportional to $\beta_{1x}$ and $\beta_{1y}$ account for polarization mode dispersion (PMD) effects \cite{agrawal2000nonlinear} characterized by a time delay $\Delta T\equiv L|v^{-1}_{gx}-v^{-1}_{gy}|=L|\beta_{1x}-\beta_{1y}|$ between the two polarization modes propagating along the $x$ and $y$ axes with different group velocities $v_{gx}$ and $v_{gy}$ respectively, $\beta_{2x}$ and $\beta_{2y}$ are the group velocity dispersion (GVD) values \cite{suzuki2001optical} of the two orthogonal polarization modes respectively, and $\alpha_x$ and $\alpha_y$ are the absorption coefficients which account for polarization-dependent loss (PDL) effects. For standard telecom fibers, the absorption coefficients are extremely small, and can hence be safely neglected. In general, the difference between the two propagation constants $\Delta\beta$ in polarization-maintaining fibers is a small quantity compared to the values of $\beta_{0x}$ and $\beta_{0y}$. To first approximation, we may replace the transverse modal fields $F_x$ and $F_y$ by the transverse distribution in Eqs.\:\eqref{TransverseDistribution1} and \eqref{TransverseDistribution2} of the corresponding fiber without shape variation in its core cross-section or intentional asymmetry stress distribution, which means that we may replace $\gamma_x$ and $\gamma_y$ by the nonlinearity parameter $\gamma$ appearing in Eq.\:\eqref{GammaFormula}, so that we obtain the following coupled-mode equations
\begin{subequations}
\begin{align}
&i\frac{\partial A_x}{\partial z}+i\beta_{1x}\frac{\partial A_x}{\partial t}-\frac{\beta_{2x}}{2}\frac{\partial^2 A_x}{\partial t^2} \nonumber\\ 
&+\gamma\left(|A_x|^2A_x+\frac{2}{3}|A_y|^2A_x+\frac{1}{3}A_y^2A_x^*e^{-2i\Delta\beta z}\right)=0,\\
&i\frac{\partial A_y}{\partial z}+i\beta_{1y}\frac{\partial A_y}{\partial t}-\frac{\beta_{2y}}{2}\frac{\partial^2 A_y}{\partial t^2} \nonumber\\ 
&+\gamma\left(|A_y|^2A_y+\frac{2}{3}|A_x|^2A_y+\frac{1}{3}A_x^2A_y^*e^{2i\Delta\beta z}\right)=0.
\end{align}
\end{subequations}
The significance of dispersion and nonlinear effects depends on the characteristics of the pulse. For the case where the fiber length $L$ is much shorter than the dispersion length $L_D\equiv T_0^2/|\beta_2|$ \cite{agrawal2000nonlinear}, dispersion effects play a small role and may be neglected, where $T_0$ is the width of the input pulse, and $\beta_2$ can be selected as the GVD value for a fiber without shape variation in its core cross-section or intentional asymmetry stress distribution. For standard telecom fibers, dispersion effects are negligible for $L<50$ km, if the pulse width $T_0>100$ ps \cite{agrawal2000nonlinear}. In particular, for CW radiation which has a stable output power over an interval of seconds or longer, we may safely neglect dispersion effects, and obtain the simplified coupled-mode equations
\begin{subequations}
\begin{align}\label{CoupleModeEquation1}
&i\frac{d A_x}{d z}=-\gamma\left(|A_x|^2A_x+\frac{2}{3}|A_y|^2A_x+\frac{1}{3}A_y^2A_x^*e^{-2i\Delta\beta z}\right),\\
&i\frac{d A_y}{d z}=-\gamma\left(|A_y|^2A_y+\frac{2}{3}|A_x|^2A_y+\frac{1}{3}A_x^2A_y^*e^{2i\Delta\beta z}\right).\label{CoupleModeEquation2}
\end{align}
\end{subequations}
For the case where the fiber length is much longer than the beat length $L_B=2\pi/|\Delta\beta|$, the four-wave-mixing terms in Eqs.\:\eqref{CoupleModeEquation1} and \eqref{CoupleModeEquation2} change signs often so that their contributions average out to be zero. Hence, for high-birefringent fibers with a typical beat length of $L_B\approx 1$ cm, the four-wave-mixing terms can be neglected. Under such an approximation, Eqs.\:\eqref{CoupleModeEquation1} and \eqref{CoupleModeEquation2} may be solved by the transformation $A_k\equiv\sqrt{P_k}e^{i\phi_k}$ \cite{agrawal2000nonlinear}, where $P_x$ and $P_y$ are the powers of the two polarization components, which are constants of motion. A straightforward calculation yields $\phi_x=\gamma(P_x+\frac{2}{3}P_y)z$ and $\phi_y=\gamma(P_y+\frac{2}{3}P_x)z$, which results in a phase difference $\Delta\phi_{NL}\equiv \phi_x-\phi_y =\frac{\gamma}{3}(P_x-P_y)z$ between the two polarization components. On the other hand, for low-birefringent fibers with a typical beat length $L_B\approx 1$ m, the four-wave-mixing terms should be retained. Another characteristic length scale is the nonlinear length $L_{NL}\equiv 1/(\gamma P_0)$ \cite{agrawal2000nonlinear}, where $P_0$ is the peak power of the incident pulse. For the case where the fiber length $L$ is much shorter than the nonlinear length $L_{NL}$, nonlinear effects play little role and may be neglected. For standard telecom fibers, nonlinear effects are negligible for $L < 50$ km, if $P_0<1$ mW \cite{agrawal2000nonlinear}.

\section{Polarization effects in anisotropic fibers}\label{3}
Expanding upon the previous discussion of electromagnetic field propagation in nonlinear single-mode fibers composed of isotropic media and birefringence effects in nonlinear weakly guiding polarization-maintaining fibers, this section focuses on the propagation characteristics of weakly guiding polarization-maintaining fibers composed of anisotropic media. To establish a foundation for studying optical analogues of quantum phase transitions in interacting quantum systems, we derive sets of coupled-mode equations. Rather than revisiting the details of the derivation, we directly present the final result, showing that the slowly varying wave amplitudes $A_j(z,t)$ satisfy the following coupled-mode equations
\begin{align}\label{AnisotropicNonlinearSchrodinger}
&i\left(\frac{\partial A_j}{\partial z}+\beta_{1j}\frac{\partial A_j}{\partial t}+\frac{\alpha_j}{2}A_j\right)-\beta_{2j}\frac{\partial^2A_j}{\partial t^2}+a_j|A_j|^2A_j\nonumber\\
&+b_j\left(2|A_k|^2A_j+A_k^2A_j^*e^{2i(\beta_{0k}-\beta_{0j})z}\right)+c_j\left(2|A_j|^2A_k e^{i(\beta_{0k}-\beta_{0j})z}\right.\nonumber\\
&\left.+A_j^2A_k^*e^{i(\beta_{0j}-\beta_{0k})z}\right)+d_j|A_k|^2A_ke^{i(\beta_{0k}-\beta_{0j})z}=0,
\end{align}
where $\alpha_j\equiv \frac{k_0^2}{2\beta_{0j}}\mbox{Im}\tilde{\chi}_{jj}^{(1)}(\omega_0)$ accounts for fiber losses, and $\beta_{1j}$ and $\beta_{2j}$ are the first and second order derivatives of $\beta_j$ evaluated at the carrier frequency $\omega_0$ respectively.  $\beta_{1j}=v_{gj}^{-1}$ is the inverse of the modal group velocity, and any difference between $\beta_{1x}$ and $\beta_{1y}$ leads to modal dispersion effects. $\beta_{2j}=\partial v_{gj}^{-1}/\partial \omega$ is the modal group velocity dispersion parameter, which is responsible for dispersive pulse broadening for the $j$-th polarization component. The remaining parameters $a_j$, $b_j$, $c_j$ and $d_j$ are defined by
\begin{align}
&a_j\equiv \gamma_{jjjj}, b_j\equiv\frac{1}{3}(\gamma_{jjkk}+\gamma_{jkjk}+\gamma_{jkkj})\nonumber\\
&c_j\equiv\frac{1}{3} (\gamma_{jjkj}+\gamma_{jkjj}+\gamma_{jjjk}), d_j\equiv \gamma_{jkkk},
\end{align}
where $\gamma_{jklm}$ are the nonlinearity parameters defined by
\begin{equation}\label{NonlinearAnisotropicParameters}
\gamma_{jklm}\equiv\frac{3k_0^2}{8\beta_0}\chi_{jklm}^{(3)}\int_{-\infty}^{\infty}\int_{-\infty}^{\infty}|F|^4 dxdy \Big/\int_{-\infty}^{\infty}\int_{-\infty}^{\infty}|F|^2 dxdy. 
\end{equation}
In the following, we always assume that the slow axis is along the $x$ direction, so that $v_{gx}<v_{gy}$ and $\beta_{0x}>\beta_{0y}$. It is convenient to introduce the transformation $u_x=A_xe^{i\Delta\beta z/2}$ and $u_y=A_ye^{-i\Delta\beta z/2}$. Under such a transformation, the coupled-mode equations can be simplified as
\begin{align}\label{RotatedCMEs}
&i\left(\frac{\partial u_j}{\partial z}+\beta_{1j}\frac{\partial u_j}{\partial t}\right)+\frac{\xi_j}{2}u_j-\frac{\beta_{2j}}{2}\frac{\partial^2 u_j}{\partial t^2}+a_j|u_j|^2u_j\nonumber\\
+&b_j\left(2|u_k|^2u_j+u_k^2u_j^*\right)+c_j\left(2|u_j|^2u_k+u_j^2u_k^*\right)+d_j|u_k|^2u_k=0,
\end{align}
where $\xi_j\equiv \beta_{0j}-\beta_{0k}+i\alpha_j$ is a complex parameter. Notice that the exponential oscillating terms in Eq.\:\eqref{AnisotropicNonlinearSchrodinger} are now replaced by the linear terms proportional to $u_j$ in Eq.\:\eqref{RotatedCMEs}.

The preceding discussion does not depend on specific material properties. However, accurately describing the interaction between the two polarization components in anisotropic fibers requires an understanding of the symmetry properties of nonlinear susceptibilities. In the following, we examine the coupled-mode equations for anisotropic fibers with different crystal symmetries. Crystal symmetries are categorized into ten cyclic three-dimensional point groups—$1$, $2$, $3$, $4$, $6$, $\bar{1}$, $m$, $\bar{3}$, $\bar{4}$, and $\bar{6}$—which combine to form thirty-two crystal point groups. These are further grouped into seven distinct crystal systems \cite{bradley2010mathematical}.

\subsubsection{Isotropic Fibers}
For isotropic fibers, the third-order nonlinear susceptibility tensor $\chi^{(3)}$ has eight nonzero elements, of which only three are independent, and its elements obey \cite{boyd2003nonlinear}
\begin{equation}\label{isotropic}
\begin{array}{l}
   \chi^{(3)}_{xxyy} = \chi^{(3)}_{yyxx},\:\chi^{(3)}_{xyxy} = \chi^{(3)}_{yxyx},\:\chi^{(3)}_{xyyx} = \chi^{(3)}_{yxxy},\\
   \chi^{(3)}_{xxxx} = \chi^{(3)}_{yyyy}  = \chi^{(3)}_{xxyy} + \chi^{(3)}_{xyxy} + \chi^{(3)}_{xyyx}.
\end{array}   
\end{equation}
Hence, we have $a_j=3b_j$, $c_j=d_j=0$ and $a_x=a_y=\gamma$ in Eq.\:\eqref{RotatedCMEs}, and $u_j$ satisfies a set of simplified coupled-mode equations
\begin{align}\label{isotropicEquation}
&i\left(\frac{\partial u_j}{\partial z}+\beta_{1j}\frac{\partial u_j}{\partial t}\right)+\frac{\xi_j}{2}u_j-\frac{\beta_{2j}}{2}\frac{\partial^2 u_j}{\partial t^2}\nonumber\\
&+\gamma\left(|u_j|^2u_j+\frac{2}{3}|u_k|^2u_j+\frac{1}{3}u_k^2u_j^*\right)=0.
\end{align}
For low-birefringent fibers, we may employ the approximations $\beta_{1x}\approx\beta_{2j}\approx\beta_1$ and $\beta_{2x}\approx\beta_{2y}\approx\beta_2$ in Eq.\:\eqref{isotropicEquation}, so that we may introduce a retarded frame by making the transformation $\tau\equiv t-\beta_1z$. In the retarded frame, the coupled-mode equations become
\begin{equation}\label{RetardedisotropicEquation}
i\frac{\partial u_j}{\partial z}=-\frac{\xi_j}{2}u_j+\frac{\beta_2}{2}\frac{\partial^2 u_j}{\partial \tau^2}-\gamma\left(|u_j|^2u_j+\frac{2}{3}|u_k|^2u_j+\frac{1}{3}u_k^2u_j^*\right).
\end{equation}
For fibers with negligible losses, we may further employ $\alpha_j\approx 0$ in Eq.\:\eqref{RetardedisotropicEquation}, so that it can be written in the form of the nonlinear Schr\"{o}dinger equation as
\begin{equation}
i\frac{\partial u_j}{\partial z}=\frac{\delta H}{\delta u_j^*},
-i\frac{\partial u_j^*}{\partial z}=\frac{\delta H}{\delta u_j},
\end{equation}
where the variational derivatives are defined by
\begin{subequations}
\begin{align}
\frac{\delta u_i(t)}{\delta u_j(t')}&=\frac{\delta u^*_i(t)}{\delta u^*_j(t')}=\delta_{ij}\delta(t-t'),\\
\frac{\delta u_i(t)}{\delta u^*_j(t')}&=\frac{\delta u^*_i(t)}{\delta u_j(t')}=0,
\end{align}
\end{subequations}
and the Hamiltonian has the form
\begin{align}
&H=-\frac{1}{2}\int d\tau\left\{\beta_2\left(\left|\partial_\tau u_x\right|^2+\left|\partial_\tau u_y\right|^2\right)+\Delta\beta(|u_x|^2-|u_y|^2)\right.\nonumber\\
&+\left.\frac{2\gamma}{3}(|u_x|^2+|u_y|^2)^2+\frac{\gamma}{3}(|u_x|^2-|u_y|^2)^2+\frac{\gamma}{3}(u_x^*u_y+u_y^*u_x)^2\right\},
\end{align}
which may also be expressed as
\begin{align}\label{IsotropicHamiltonian}
&H=-\frac{1}{2}\int d\tau\left\{\beta_2\left(\left|\partial_\tau u_x\right|^2+\left|\partial_\tau u_y\right|^2\right)\right.\nonumber\\
&+\left.\Delta\beta S_z+\frac{2\gamma}{3}S^2+\frac{\gamma}{3}(S_z^2+S_x^2)\right\},
\end{align}
where $S\equiv |u_x|^2+|u_y|^2$, $S_x\equiv u_x^*u_y+u_y^*u_x$, $S_y \equiv i\left(u_x^*\,u_y - u_y^*\,u_x\right)$ and $S_z\equiv |u_x|^2-|u_y|^2$. This formulation corresponds to the well-known Hopf map, which establishes a mapping from the three-dimensional sphere $S^3$ onto a two-dimensional sphere $S^2$ via a fiber bundle structure, where where each point on $S^2$ corresponds to an entire circle in $S^3$. Since $S^2=S_x^2+S_y^2+S_z^2$ is conserved, the Hamiltonian in Eq.\:\eqref{IsotropicHamiltonian} comprises a rotor-like term proportional to \( S_z \) and another Euler-top like term proportional to \( S_y^2 \). This formulation serves as an optical analogue of the standard LMG model, with \( \Delta\beta \), the difference in the phase accumulation rates of the polarization components, acting as the applied external field and \( \gamma \) quantifying the interaction strength. The rotor-like term steers the evolution of the polarization state by setting a relative phase difference between the polarization components. Thus, the standard LMG model—and its associated quantum phase transitions—has a direct analogue in the nonlinear polarization dynamics and classical bifurcation observed in an isotropic optical fiber. Since the dynamical behavior of the standard LMG model is well-established, we will not discuss its details here and instead move directly on to exploring more complicated symmetry classes in fiber materials.

\subsubsection{Cubic Symmetry}
We now discuss anisotropic fibers with cubic symmetry. The cubic crystal system has five point groups, $23$, $m3$, $432$, $\bar{4}3m$ and $m3m$ \cite{bradley2010mathematical}. For the two classes $23$ and $m3$, the third-order nonlinear susceptibility tensor $\chi^{(3)}$ has eight nonzero elements and seven of them are independent. Those independent elements are given by $\chi^{(3)}_{xxxx} = \chi^{(3)}_{yyyy},\:\chi^{(3)}_{xxyy},\:\chi^{(3)}_{yyxx},\:\chi^{(3)}_{xyxy},\:\chi^{(3)}_{yxyx},\:\chi^{(3)}_{xyyx},\:\chi^{(3)}_{yxxy}$ \cite{boyd2003nonlinear}. Hence, we have $a_x=a_y=a$ and $c_j=d_j=0$ in \eqref{RotatedCMEs}, and $u_j$ satisfies the following coupled-mode equations in the retarded frame
\begin{equation}\label{23andm3}
i\frac{\partial u_j}{\partial z}=-\frac{\xi_j}{2}u_j+\frac{\beta_2}{2}\frac{\partial^2 u_j}{\partial \tau^2}-a|u_j|^2u_j-b_j(2|u_k|^2u_j+u_k^2u_j^*).
\end{equation}
Eq.\:\eqref{23andm3} can be expressed in the form of the nonlinear Schr\"{o}dinger equation only when $b_x=b_y$. The details of the arguments are provided in App.\:\ref{B}. For the remaining three classes $432$, $\bar{4}3m$ and $m3m$, the third-order nonlinear susceptibility tensor $\chi^{(3)}$ has eight nonzero elements, of which only four are independent, and its elements obey \cite{boyd2003nonlinear}
\begin{equation}\label{cubicminor}
\begin{array}{l}
\chi^{(3)}_{xxxx} = \chi^{(3)}_{yyyy},\:\chi^{(3)}_{xxyy} = \chi^{(3)}_{yyxx},\\
\chi^{(3)}_{xyxy} = \chi^{(3)}_{yxyx},\:\chi^{(3)}_{xyyx} = \chi^{(3)}_{yxxy}.
\end{array}
\end{equation}
Hence, we have $a_x=a_y=a$, $b_x=b_y=b$ and $c_j=d_j=0$ in Eq.\:\eqref{RotatedCMEs}, so that $u_j$ satisfies the following coupled-mode equations in the retarded frame
\begin{equation}\label{cubicminorequations}
i\frac{\partial u_j}{\partial z}=-\frac{\xi_j}{2}u_j+\frac{\beta_2}{2}\frac{\partial^2 u_j}{\partial \tau^2}-a|u_j|^2u_j-b(2|u_k|^2u_j+u_k^2u_j^*).
\end{equation}
For fibers with negligible losses, Eq.\:\eqref{cubicminorequations} can be written in the form of the nonlinear Schr\"{o}dinger equation, and the Hamiltonian is given by
\begin{align}
&H=-\frac{1}{2}\int d\tau\left\{\beta_2\left(\left|\partial_\tau u_x\right|^2+\left|\partial_\tau u_y\right|^2\right)+\Delta\beta (|u_x|^2-|u_y|^2)\right.\nonumber\\
&\left.+c_0(|u_x|^2+|u_y|^2)^2+c_z(|u_x|^2-|u_y|^2)^2+c_x(u_x^*u_y+u_y^*u_x)^2\right\},
\end{align}
which may also be expressed as
\begin{align}\label{cubicHamiltonian}
&H=-\frac{1}{2}\int d\tau\left\{\beta_2\left(\left|\partial_\tau u_x\right|^2+\left|\partial_\tau u_y\right|^2\right)\right.\nonumber\\
&+\left.\Delta\beta S_z+c_0S_0^2+c_zS_z^2+c_xS_x^2\right\},
\end{align}
where the nonlinearity parameters $c_0$, $c_z$ and $c_x$ are defined by $c_0\equiv \frac{1}{2}(a+b)$, $c_z\equiv \frac{1}{2}(a-b)$ and $c_x\equiv b$. For the specific case where $a=3b=\gamma$, the Hamiltonian Eq.\:\eqref{cubicHamiltonian} for cubic media reduces to the Hamiltonian Eq.\:\eqref{IsotropicHamiltonian} for isotropic media. The Hamiltonian in Eq.\,\eqref{cubicHamiltonian} shares the form of the standard LMG model. In this model, a quantum phase transition---or its corresponding classical bifurcation---occurs when the rotor-like term becomes comparable to the Euler-top-like term. However, the top-like term by itself does not induce a quantum phase transition. Interestingly, the quadratic terms consist of two parts. One part involves $S_y^2$, which corresponds to single-axis squeezing, and the other part involves $S_z^2 - S_x^2$, which corresponds to two-axis squeezing. When applied to the lowest weight state, these squeezing terms produce one-axis and two-axis squeezed states, respectively.

\subsubsection{Hexagonal Symmetry}
We now discuss anisotropic fibers with hexagonal symmetry. The hexagonal crystal system has seven point groups, $6$, $\bar{6}$, $6/m$, $622$, $6mm$, $\bar{6}2m$ and $6/mmm$ \cite{bradley2010mathematical}. For the three crystal classes $6$, $\bar{6}$ and $6/m$, the third-order nonlinear susceptibility tensor $\chi^{(3)}$ has sixteen nonzero elements, of which only six are independent, and its elements obey \cite{boyd2003nonlinear}
\begin{equation}\label{hexagonal}
  \begin{array}{l}
  \left\{
   \begin{array}{c}
   \chi^{(3)}_{xxyy} = \chi^{(3)}_{yyxx},\\
   \chi^{(3)}_{xyxy} = \chi^{(3)}_{yxyx},\\
  \chi^{(3)}_{xyyx} = \chi^{(3)}_{yxxy},
   \end{array}
  \right.
  \left\{
   \begin{array}{c}
   \chi^{(3)}_{yyxy} = -\chi^{(3)}_{xxyx},  \\
   \chi^{(3)}_{yxyy} = -\chi^{(3)}_{xyxx},  \\
   \chi^{(3)}_{xyyy} = -\chi^{(3)}_{yxxx},  
   \end{array}
  \right.  \\ 
\chi^{(3)}_{xxxx}=\chi^{(3)}_{yyyy}=\chi^{(3)}_{xxyy}+\chi^{(3)}_{xyyx}+\chi^{(3)}_{xyxy},\\ 
\chi^{(3)}_{xxxy}=-\chi^{(3)}_{yyyx}=\chi^{(3)}_{yyxy}+\chi^{(3)}_{yxyy}+\chi^{(3)}_{xyyy}.
 \end{array}
\end{equation}
Hence, the parameters $a_j$, $b_j$, $c_j$ and $d_j$ in Eq.\:\eqref{RotatedCMEs} satisfy
\begin{equation}
  \begin{array}{l}
  \left\{
   \begin{array}{c}
   a_x = a_y = a,\\
   b_x = b_y = b,
   \end{array}
  \right.
  \left\{
   \begin{array}{c}
   c_x = -c_y = c,  \\
   d_x = -d_y = d,  
   \end{array}
  \right.  \\ 
 \end{array}
 a=3b\:\;\mbox{and}\:\;d=3c,
\end{equation}
so that $u_j$ obeys the following coupled-mode equations in the retarded frame
\begin{align}\label{6bar66m}
i\frac{\partial u_j}{\partial z}=&-\frac{\xi_j}{2}u_j+\frac{\beta_2}{2}\frac{\partial^2 u_j}{\partial \tau^2}-\frac{a}{3}\left(3|u_j|^2u_j+2|u_k|^2u_j+u_k^2u_j^*\right)\nonumber\\
&\mp\frac{d}{3}\left(3|u_k|^2u_k+2|u_j|^2u_k+u_j^2u_k^*\right),
\end{align}
where the minus and plus signs correspond to \(j=x\) and \(j=y\), respectively. Equation~\eqref{6bar66m} can be written in the form of the nonlinear Schrödinger equation only when \(d=-c=0\). The details of the arguments are provided in App.\:\ref{B}.

For the crystal classes \(622\), \(6mm\), \(\bar{6}2m\), and \(6/mmm\), the third-order nonlinear susceptibility tensor \(\chi^{(3)}\) has sixteen nonzero elements, with only three of them being independent. The nonzero elements of \(\chi^{(3)}\) obey Eq.~\eqref{isotropic}, which leads to a set of coupled-mode equations in the form of Eq.~\eqref{isotropicEquation}. That is, for fiber materials with hexagonal symmetry, either they cannot correspond to a generalized LMG Hamiltonian, or their nonlinear polarization dynamics reduce to those described by isotropic fibers.

\subsubsection{Trigonal Symmetry}

We now turn our attention to anisotropic fibers exhibiting trigonal symmetry. The trigonal crystal system encompasses five point groups: \(3\), \(\bar{3}\), \(32\), \(3m\), and \(\bar{3}m\) \cite{bradley2010mathematical}. In the crystal classes \(3\) and \(\bar{3}\), the third-order nonlinear susceptibility tensor \(\chi^{(3)}\) contains sixteen nonzero elements, of which only six are independent. These nonzero elements satisfy Eq.~\eqref{hexagonal}, leading to a set of coupled-mode equations in the form of Eq.~\eqref{6bar66m}.

On the other hand, for the crystal classes \(3m\), \(\bar{3}m\), and \(32\), the tensor \(\chi^{(3)}\) exhibits eight nonzero elements, with only four being independent. In this instance, the nonzero elements obey Eq.~\eqref{isotropic}, giving rise to coupled-mode equations described by Eq.~\eqref{isotropicEquation}. In both cases, the nonlinear polarization associated with trigonal symmetry either cannot be precisely mapped onto a generalized LMG Hamiltonian or reduces to the same form as that found in isotropic fibers. In the following, we focus on tetragonal symmetry, where nontrivial nonlinear polarization dynamics emerge.

\subsubsection{Tetragonal Symmetry}

The tetragonal crystal system has seven point groups, $4$, $\bar{4}$, $4/m$, $422$, $4mm$, $\bar{4}2m$ and $4/mmm$ \cite{bradley2010mathematical}. For the three crystal classes $4$, $\bar{4}$ and $4/m$, the third-order nonlinear susceptibility tensor $\chi^{(3)}$ has sixteen nonzero elements, of which only eight are independent, and its elements obey \cite{boyd2003nonlinear}
\begin{equation}
  \begin{array}{l}
  \left\{
   \begin{array}{c}
   \chi^{(3)}_{xxyy} = \chi^{(3)}_{yyxx},\\
   \chi^{(3)}_{xyxy} = \chi^{(3)}_{yxyx},\\
   \chi^{(3)}_{xyyx} = \chi^{(3)}_{yxxy},\\
   \chi^{(3)}_{xxxx}=\chi^{(3)}_{yyyy},
   \end{array}
  \right.
  \left\{
   \begin{array}{c}
   \chi^{(3)}_{yyxy} = -\chi^{(3)}_{xxyx},  \\
   \chi^{(3)}_{yxyy} = -\chi^{(3)}_{xyxx},  \\
   \chi^{(3)}_{xyyy} = -\chi^{(3)}_{yxxx},  \\
   \chi^{(3)}_{xxxy}=-\chi^{(3)}_{yyyx}.
   \end{array}
  \right.  \\ 
  \end{array}
\end{equation}
Hence, from the above symmetry requirements, the parameters $a_j$, $b_j$, $c_j$ and $d_j$ in Eq.\:\eqref{RotatedCMEs} satisfy
\begin{equation}
  \begin{array}{l}
  \left\{
   \begin{array}{c}
   a_x = a_y = a,\\
   b_x = b_y = b,
   \end{array}
  \right.
  \left\{
   \begin{array}{c}
   c_x = -c_y = c,  \\
   d_x = -d_y = d,
   \end{array}
  \right.  \\ 
 \end{array}
\end{equation}
so that $u_j$ obeys the following coupled-mode equations in the retarded frame
\begin{align}\label{4bar44m}
i\frac{\partial u_j}{\partial z}=&-\frac{\xi_j}{2}u_j+\frac{\beta_2}{2}\frac{\partial^2 u_j}{\partial \tau^2}-a|u_j|^2u_j-b\left(2|u_k|^2u_j+u_k^2u_j^*\right)\nonumber\\
&\mp c\left(2|u_j|^2u_k+u_j^2u_k^*\right)\mp d|u_k|^2u_k,
\end{align}
where the minus and plus signs correspond to $j=x$ and $j=y$ respectively. Eq.\:\eqref{4bar44m} is integrable only when the condition $d=-c$ is fulfilled. This condition leads to
\begin{align}\label{IntegrableNSE}
i\frac{\partial u_j}{\partial z}=&-\frac{\xi_j}{2}u_j+\frac{\beta_2}{2}\frac{\partial^2 u_j}{\partial \tau^2}-a|u_j|^2u_j-b\left(2|u_k|^2u_j+u_k^2u_j^*\right)\nonumber\\
&\mp c\left(2|u_j|^2u_k+u_j^2u_k^*-|u_k|^2u_k\right).
\end{align}
For fibers with negligible losses, Eq.\:\eqref{IntegrableNSE} can be written in the form of the nonlinear Schr\"{o}dinger equation, and the Hamiltonian is given by
\begin{align}
H&=-\frac{1}{2}\int d\tau\left\{\beta_2\left(\left|\partial_\tau u_x\right|^2+\left|\partial_\tau u_y\right|^2\right)\right.\nonumber\\
&\left.+\Delta\beta (|u_x|^2-|u_y|^2)+c_0(|u_x|^2+|u_y|^2)^2\right.\nonumber\\
&\left.+c_z(|u_x|^2-|u_y|^2)^2+c_x(u_x^*u_y+u_y^*u_x)^2\right.\nonumber\\
&\left.+2c(|u_x|^2-|u_y|^2)(u_x^*u_y+u_y^*u_x)\right\},
\end{align}
which may also be expressed as
\begin{align}\label{TetragonalHamiltonian}
H=&-\frac{1}{2}\int d\tau\left\{\beta_2\left(\left|\partial_\tau u_x\right|^2+\left|\partial_\tau u_y\right|^2\right)+\Delta\beta S_z\right.\nonumber\\
&\left.+c_0S^2+c_zS_z^2+c_xS_x^2+2cS_zS_x\right\},
\end{align}
where $c_0\equiv \frac{1}{2}(a+b)$, $c_z\equiv \frac{1}{2}(a-b)$ and $c_x\equiv b$, as in Eq.\:\eqref{cubicHamiltonian}. For the four crystal classes $422$, $4mm$, $\bar{4}2m$ and $4/mmm$, the third-order nonlinear susceptibility tensor $\chi^{(3)}$ has eight nonzero elements and only four of them are independent. The nonzero elements of the third-order nonlinear susceptibility obey Eq.\:\eqref{cubicminor}, which leads to a set of coupled-mode equations in the form of Eq.\:\eqref{cubicminorequations}.

From the above, it is evident that tetragonal symmetry offers the simplest nontrivial scenario. In Eq.\:\eqref{TetragonalHamiltonian}, the Hamiltonian features a quadratic crossing term, \( S_zS_x \), which is absent from the conventional LMG model. In addition to the standard single-axis squeezing term, \( S^2 - S_y^2 \), and the two-axis squeezing term, \( S_z^2 - S_x^2 \), the appearance of this unconventional squeezing term, \( S_zS_x \), gives rise to a novel quantum phase transition in the generalized LMG model.

This transition manifests as an excited-state quantum phase transition. At the transition point, not only do the ground state and the lowest excited state cross, but a series of excited states also intersect. This scenario contrasts with cases where numerous energy levels converge at a single critical point; instead, beyond the critical point, a sequence of crossing points emerges. Moreover, the geometric phase acquired by the wave function as it traverses the parameter space—generated by a singular gauge field—is not characterized by a monopole or a non-Abelian monopole. Rather, it takes the form of a conical cosmological singularity in a de Sitter space.

In parallel, the quantum phase transition described above corresponds to a classical bifurcation in the nonlinear polarization dynamics on the Poincar\'{e} sphere. When the conventional one-axis and two-axis squeezing terms dominate, the nonlinear polarization dynamics are analogous to a rigid Euler top with positive effective mass. Conversely, when the unconventional \( S_zS_x \) term prevails, the dynamics mirror those of a rigid Euler top with negative effective mass.

In the following section, we describe the dynamics of the classical optical analogue of the generalized LMG model. The primary focus of this paper is a comprehensive study of the classical bifurcation in the nonlinear polarization dynamics along an optical fiber, corresponding to the large-spin limit of the generalized LMG model. We defer the discussion of the spectral properties of the excited-state quantum phase transition and the associated squeezing phenomena to a forthcoming paper.

\subsubsection{Monoclinic and Triclinic Symmetries}

Last but not least, the remaining crystal systems are: the monoclinic crystal system, which has three point groups $2$, $m$ and $2/m$; the orthorhombic crystal system, which has three point groups $222$, $mm2$ and $mmm$; and the triclinic crystal system, which has two point groups $1$ and $\bar{1}$ \cite{bradley2010mathematical}. The coupled-mode equations Eq.\:\eqref{RotatedCMEs} for anisotropic fibers with these crystal symmetries are integrable only when the conditions $b_x=b_y$, $d_y=c_x$ and $d_x=c_y$ are fulfilled. The details of the arguments are provided in App.\:\ref{B}. Hence the resulting dynamics are in general dissipative. Since the scope of this work is restricted to geometric phases for integrable systems, the physics of the dissipative dynamics mentioned above will be left to future research.

\section{Properties of the Classical Spin Model}\label{5}
In this section, we explore key properties of the classical spin model derived from nonlinear polarization dynamics, incorporating a non-conventional squeezing term. We specifically examine the scenario where the linear rotor-like term is absent, focusing instead on the quadratic components. These include conventional one-axis and two-axis squeezing terms, as well as the unconventional squeezing term
\begin{equation}\label{ClassicalSpinHamiltonian}
H=\frac{\alpha}{2}S_x^2+\beta S_xS_y+\frac{\gamma}{2}S_y^2.
\end{equation}
Here $\alpha$, $\beta$ and $\gamma$ are some real parameters, and $S_x$, $S_y$ and $S_z$ are the components of the spin which satisfy the standard Poisson bracket $\{S_i,S_j\}=\epsilon_{ijk}S_k$. From the Hamiltonian Eq.\:\eqref{ClassicalSpinHamiltonian}, the spin's equations of motion for fixed parameters are derived using Hamilton's equation, expressed as $\dot{S}_i=\{S_i,H\}$. These equations are explicitly formulated as
\begin{subequations}
\begin{align}\label{EOM}
\dot{S}_x&=(\beta S_x+\gamma S_y)S_z, \\
\dot{S}_y&=-(\beta S_y+\alpha S_x)S_z, \\
\dot{S}_z&=(\alpha-\gamma)S_xS_y+\beta(S_y^2-S_x^2).\label{EOM3}
\end{align}
\end{subequations}
The classical spin system is integrable as we have two integrals of motion, which can be derived from the laws of conservation of energy and angular momentum
\begin{subequations}
\begin{align}
2H&=\alpha S_x^2+2\beta S_xS_y+\gamma S_y^2,\\
S^2&=S_x^2+S_y^2+S_z^2.
\end{align}
\end{subequations}
This equi-energy curve can be regarded as an algebraic variety, characterized as the common locus of the two polynomial equations presented above. In the following, we set $S=1$. To bring the Hamiltonian Eq.\:\eqref{ClassicalSpinHamiltonian} into a simpler form, we perform a coordinate transformation 
\begin{equation}
M_x=\cos\theta S_x -\sin\theta S_y, M_y=\sin\theta S_x + \cos\theta S_y, M_z=S_z,
\end{equation}
where $\tan2\theta\equiv 2\beta/(\alpha-\gamma)$. After the coordinate transformation, the Hamiltonian from Eq.\:\eqref{ClassicalSpinHamiltonian} becomes
\begin{equation}
H=\frac{\alpha+\gamma}{2}\left(\frac{M_x^2+M_y^2}{2}\right)+\sqrt{\left(\frac{\alpha-\gamma}{2}\right)^2+\beta^2}\left(\frac{M_x^2-M_y^2}{2}\right),
\end{equation}
which can be viewed as the Hamiltonian of an asymmetric Euler top \cite{landau1976mechanics}
\begin{equation}
H=\frac{M_x^2}{2I_x}+\frac{M_y^2}{2I_y},
\end{equation}
\begin{figure}[htbp]
\centering
\fbox{\includegraphics[width=0.98\linewidth]{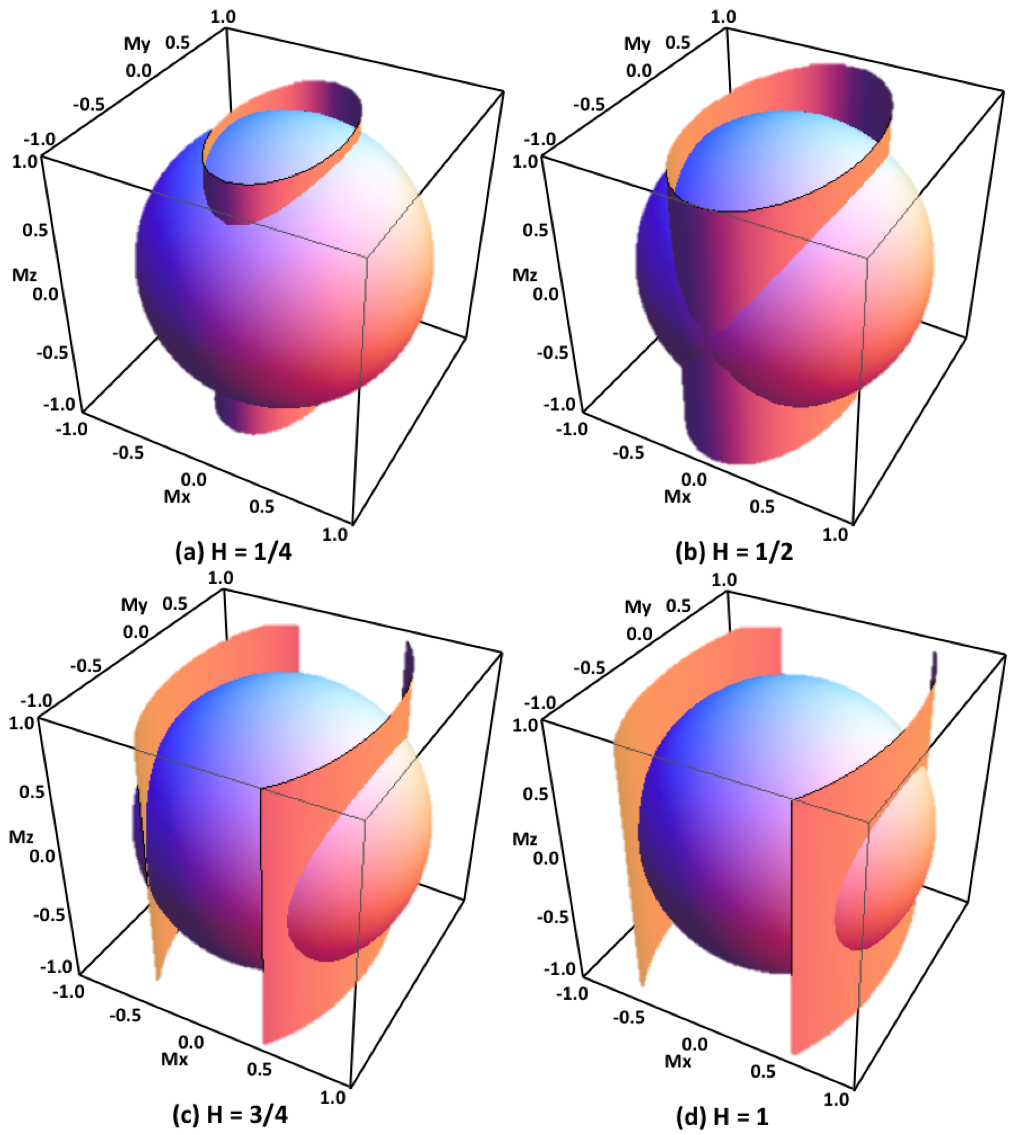}}
\caption{Intersection of the elliptic cylinder $M_x^2/I_x+M_y^2/I_y=2H$ with the unit sphere $M_x^2+M_y^2+M_z^2=1$ at different energies $H$. Here we set $\alpha=\gamma=2$ and $\beta=1$, so that $I_x=1/3$ and $I_y=1$. There are intersections between the elliptic cylinder and the unit sphere if and only if the condition $0<H<3/2$ is fulfilled.}
\label{fig:CylinderAndSphere}
\end{figure}
where $I_x$ and $I_y$ are the principal moments of inertia of the corresponding asymmetric Euler top defined by
\begin{subequations}
\begin{align}\label{AsymmetricEulerTop1}
\frac{1}{I_x}&\equiv \frac{1}{2}\left(\alpha+\gamma+\sqrt{\left(\alpha-\gamma\right)^2+4\beta^2}\right),\\
\frac{1}{I_y}&\equiv \frac{1}{2}\left(\alpha+\gamma-\sqrt{\left(\alpha-\gamma\right)^2+4\beta^2}\right).\label{AsymmetricEulerTop2}
\end{align}
\end{subequations}
In terms of $M_x$, $M_y$ and $M_z$, the laws of conservation of energy and angular momentum become
\begin{subequations}
\begin{align} \label{EnergyMomentum1}
2H&=M_x^2/I_x+M_y^2/I_y,\\
1&=M_x^2+M_y^2+M_z^2.\label{EnergyMomentum2}
\end{align}
\end{subequations}
From Eqs.\:\eqref{AsymmetricEulerTop1} -- \eqref{AsymmetricEulerTop2}, we note that $I_y \geq I_x$, where the equality holds if and only if $\alpha=\gamma$ and $\beta=0$. In particular, $I_y$ is positive when $\alpha\gamma>\beta^2$, which implies that the principal moments of inertia of the asymmetric top are positive only when $\alpha\gamma>\beta^2$. Hence, we have to discuss the cases $\alpha\gamma>\beta^2$ and $\alpha\gamma<\beta^2$ separately.

\subsection{The motion of the spin for $\alpha\gamma>\beta^2$}
In this subsection, we focus on the case $\alpha\gamma>\beta^2$. We notice that Eq.\:\eqref{EnergyMomentum1} and Eq.\:\eqref{EnergyMomentum2} are the equations of an elliptic cylinder with semi-axes $\sqrt{2HI_x}$, $\sqrt{2HI_y}$ and of a unit sphere respectively. The existence of an intersection is ensured by the inequalities $0<2HI_x<1$, which provide that the radius of the sphere is greater than the semi-minor-axis of the elliptic cylinder. When the vector $\mathbf{M}$ moves relative to the axes of inertia of the top, its terminus moves along the line of intersection of the two surfaces. Fig.~\ref{fig:CylinderAndSphere} shows the lines of intersection of elliptic cylinders at different energies with the unit sphere. Using Eq.\:\eqref{EOM} -- Eq.\:\eqref{EOM3}, we obtain the equations of motion of the asymmetric top in terms of $M_x$, $M_y$ and $M_z$
\begin{equation}\label{EquationsOfMotionForTop}
\dot{M}_x=-\frac{M_yM_z}{I_y},\dot{M}_y=\frac{M_xM_z}{I_x},\dot{M}_z=\left(\frac{1}{I_y}-\frac{1}{I_x}\right)M_xM_y.
\end{equation}
From Eq.\:\eqref{EquationsOfMotionForTop}, we may notice that the system contains six fixed points, which are $M_z=\pm 1$, $M_y=\pm 1$, and $M_x=\pm 1$. The motion is stable near the $M_x$ and $M_z$ axes, but is unstable in the neighborhood of the $M_y$ axis. According to Eq.\:\eqref{EnergyMomentum1} and Eq.\:\eqref{EnergyMomentum2}, we can express $M_x$ and $M_y$ in terms of $M_z$
\begin{subequations}
\begin{align}\label{Mx}
M_x^2 &= \frac{I_x}{I_y-I_x}[M_z^2-(1-2HI_y)],\\
M_y^2 &= \frac{I_y}{I_y-I_x}[(1-2HI_x)-M_z^2],\label{My}
\end{align}
\end{subequations}
where $1-2HI_y\leq M_z^2\leq 1-2HI_x$. Substitution of Eq.\:\eqref{Mx} and Eq.\:\eqref{My} into Eq.\:\eqref{EquationsOfMotionForTop} immediately yields
\begin{subequations}\label{EllipticEquationOfMotion}
\begin{align}\label{EllipticEquationOfMotion1}
\dot{M}_x &= \omega\sqrt{(2HI_x-M_x^2)[1-2HI_y+(I_y-I_x)M_x^2/I_x]},\\
\dot{M}_y &= \omega\sqrt{(2HI_y-M_y^2)[1-2HI_x-(I_y-I_x)M_y^2/I_y]},\\
\dot{M}_z &= \omega\sqrt{[(1-2HI_x)-M_z^2][M_z^2-(1-2HI_y)]},\label{EllipticEquationOfMotion3}
\end{align}
\end{subequations}
where $\omega\equiv \sqrt{\alpha\gamma-\beta^2}$. Now, we discuss the cases $2HI_y\leq 1$ and $2HI_y\geq 1$ separately. For the case $2HI_y\leq 1$, we introduce a set of new variables $u\equiv\omega t\sqrt{1-2HI_x}$, $m_x\equiv M_x/\sqrt{2HI_x}$, $m_y\equiv M_y/\sqrt{2HI_y}$, $m_z\equiv M_z/\sqrt{1-2HI_x}$, $k^2\equiv 2H(I_y-I_x)/(1-2HI_x)$, and $k'^2\equiv (1-2HI_y)/(1-2HI_x)$, so that Eq.\:\eqref{EllipticEquationOfMotion1} -- Eq.\:\eqref{EllipticEquationOfMotion3} become
\begin{subequations}\label{EllipticFunctionDn}
\begin{align}\label{EllipticFunctionDn1}
\frac{dm_x}{du} &= \sqrt{(1-m_x^2)(k'^2+k^2m_x^2)},\\
\frac{dm_y}{du} &= \sqrt{(1-m_y^2)(1-k^2m_y^2)},\\
\frac{dm_z}{du} &= \sqrt{(1-m_z^2)(m_z^2-k'^2)}.\label{EllipticFunctionDn3}
\end{align}
\end{subequations}
Eq.\:\eqref{EllipticFunctionDn1} -- Eq.\:\eqref{EllipticFunctionDn3} show that $M_x$, $M_y$ and $M_z$ are solved by the Jacobi elliptic functions of a linear function of $t$
\begin{subequations}\label{EulerTopFinalExpression}
\begin{align}\label{EulerTopFinalExpression1}
M_x &=\sqrt{2HI_x}\cn(\lambda t,k) =\cn(a,k')\cn(\lambda t,k),\\
M_y &= \sqrt{2HI_y}\sn(\lambda t,k)=\dn(a,k')\sn(\lambda t,k),\\
M_z &= \sqrt{1-2HI_x}\dn(\lambda t,k)=\sn(a,k')\dn(\lambda t,k),\label{EulerTopFinalExpression3}
\end{align}
\end{subequations}
where $\lambda\equiv \omega\sqrt{1-2HI_x}$, $a$ is a real number defined by $\cn(a,k')\equiv \sqrt{2HI_x}$, and $\sn u$, $\cn u$ and $\dn u$ are doubly periodic functions of $u$ which satisfy
\begin{subequations}
\begin{align}
\sn(u+4K,k)&=\sn(u+2iK',k)=\sn(u,k),\\
\cn(u+4K,k)&=\cn(u+2K+2iK',k)=\cn(u,k),\\
\dn(u+2K,k)&=\dn(u+4iK')=\dn(u,k),
\end{align}
\end{subequations}
where $k$ is the elliptic modulus, and $K$ and $iK'$ are the quarter periods of the elliptic functions defined by
\begin{subequations}
\begin{align}
K&\equiv \int_0^{\pi/2} \frac{d\phi}{\sqrt{1-k^2\sin^2\phi}}=\sum_{n=0}^{\infty}\left(\frac{(2n-1)!!}{(2n)!!}\right)^2k^{2n}\\
K'&\equiv \int_0^{\pi/2} \frac{d\psi}{\sqrt{1-k'^2\sin^2\psi}}=K(k').
\end{align}
\end{subequations}
The values of the Jacobi elliptic functions at the quarter periods $0$, $K$, $iK'$ and $K+iK'$ are listed in Table \ref{tab:quarter-periods}.
\begin{table}[htbp]
\centering
\caption{\bf The values of the Jacobi elliptic functions at the quarter periods}
\begin{tabular}{ccccc}
\hline
quarter periods & $0$ & $K$ & $iK'$ & $K+iK'$  \\
\hline
$\sn$ & $0$ & $1$ & $\infty$ & $1/k$ \\
$\cn$ & $1$ & $0$ & $\infty$ & $-ik'/k$ \\
$\dn$ & $1$ & $k'$ & $\infty$ & $0$ \\
\hline
\end{tabular}
  \label{tab:quarter-periods}
\end{table}
Table \ref{tab:quarter-periods} shows that each Jacobi elliptic function has a simple pole at a quarter period and a simple zero at another quarter period. In other words, each Jacobi elliptic function has only one pole and one zero within the fundamental parallelogram with vertices $u=0$, $u=2K$, $u=2iK'$ and $u=2K+2iK'$ in the complex plane. If we define the four Jacobi theta functions as a doubly infinite sum of the form
\begin{equation}\label{JacobiTheta}
\Theta_{\mu\nu}(z,\tau)\equiv \sum_{n=-\infty}^{\infty}\exp\pi i\left\{\tau\left(n+\frac{\mu}{2}\right)^2+2\left(n+\frac{\mu}{2}\right)\left(z+\frac{\nu}{2}\right)\right\},
\end{equation}
where $\mu$ and $\nu$ take values of 0 or 1, it can be shown that the Jacobi theta functions $\Theta_{\mu\nu}(z,\tau)$ converge in the entire complex plane for any $\tau$ with $\mbox{Im}\tau>0$. According to the definition Eq.\:\eqref{JacobiTheta}, the Jacobi theta functions are quasi-doubly periodic functions which obey the following functional equations
\begin{subequations}\label{JacobiThetaQuasi}
\begin{align}\label{JacobiThetaQuasi1}
\Theta_{\mu\nu}(z+1,\tau)&=(-1)^{\mu}\Theta_{\mu\nu}(z,\tau),\\
\Theta_{\mu\nu}(z+\tau,\tau)&=(-1)^{\nu}e^{-\pi i(\tau+2z)}\Theta_{\mu\nu}(z,\tau).\label{JacobiThetaQuasi2}
\end{align}
\end{subequations}
Eq.\:\eqref{JacobiThetaQuasi1} and Eq.\:\eqref{JacobiThetaQuasi2} reveal that the quotient of the Jacobi theta functions $f_{\mu\nu\rho\sigma}(z,\tau)\equiv \Theta_{\mu\nu}(z,\tau)/\Theta_{\rho\sigma}(z,\tau)$ are doubly periodic functions on the fundamental parallelogram with vertices $z=0$, $z=1$, $z=\tau$ and $z=1+\tau$ in the complex $z$ plane, which satisfy
\begin{subequations}
\begin{align}
f_{\mu\nu\rho\sigma}(z+1,\tau)&=(-1)^{\mu-\rho}f_{\mu\nu\rho\sigma}(z,\tau),\\
f_{\mu\nu\rho\sigma}(z+\tau,\tau)&=(-1)^{\nu-\sigma}f_{\mu\nu\rho\sigma}(z,\tau).
\end{align}
\end{subequations}
It can be shown that the zeros of the Jacobi theta functions are exactly at the half-periods $1/2$, $\tau/2$, $1/2+\tau/2$, and 0 within the fundamental parallelogram. In particular, we have
\begin{equation}\label{EllipticThetaZero}
\Theta_{\mu\nu}\left(\frac{1}{2}+\frac{\tau}{2}-\frac{\mu\tau}{2}-\frac{\nu}{2}\right)=0,
\end{equation}
or equivalently, $\Theta_{00}(1/2+\tau/2)=\Theta_{01}(\tau/2)=\Theta_{10}(1/2)=\Theta_{11}(0)=0$. This implies that the Jacobi elliptic functions can be represented by quotients of the Jacobi theta functions. A direct comparison between Table \ref{tab:quarter-periods} and \eqref{EllipticThetaZero} yields
\begin{subequations}
\begin{align}
\sn(u,k)&=\frac{\Theta_{01}\left(\frac{1}{2},\frac{iK'}{K}\right)\Theta_{11}\left(\frac{u}{2K},\frac{iK'}{K}\right)}{\Theta_{11}\left(\frac{1}{2},\frac{iK'}{K}\right)\Theta_{01}\left(\frac{u}{2K},\frac{iK'}{K}\right)},\\
\cn(u,k)&=\frac{\Theta_{01}\left(0,\frac{iK'}{K}\right)\Theta_{10}\left(\frac{u}{2K},\frac{iK'}{K}\right)}{\Theta_{10}\left(0,\frac{iK'}{K}\right)\Theta_{01}\left(\frac{u}{2K},\frac{iK'}{K}\right)},\\
\dn(u,k)&=\frac{\Theta_{01}\left(0,\frac{iK'}{K}\right)\Theta_{00}\left(\frac{u}{2K},\frac{iK'}{K}\right)}{\Theta_{00}\left(0,\frac{iK'}{K}\right)\Theta_{01}\left(\frac{u}{2K},\frac{iK'}{K}\right)}.
\end{align}
\end{subequations}
For the case $2HI_y\geq 1$, the elliptic modulus $k\equiv \sqrt{2H(I_y-I_x)/(1-2HI_x)}>1$, and we may apply the identities for elliptic moduli outside the interval $[0,1]$: $\sn(u,k)=k^{-1}\sn(ku,k^{-1})$, $\cn(u,k)=\dn(ku,k^{-1})$ and $\dn(u,k)=\cn(ku,k^{-1})$. According to Eq.\:\eqref{EulerTopFinalExpression1} -- Eq.\:\eqref{EulerTopFinalExpression3}, $M_x$, $M_y$ and $M_z$ may be written as
\begin{subequations}\label{FinalElliptic}
\begin{align}\label{FinalElliptic1}
M_x &= \sqrt{2HI_x}\dn(\kappa t,k_1)=\cd(ka,k'_1)\dn(\kappa t,k_1),\\
M_y &= \sqrt{I_y(1-2HI_x)/(I_y-I_x)}\sn(\kappa t,k_1)\nonumber\\
&=k_1\nd(ka,k'_1)\sn(\kappa t,k_1),\\
M_z &= \sqrt{1-2HI_x}\cn(\kappa t,k_1)= k_1\sd(ka,k'_1)\cn(\kappa t,k_1),\label{FinalElliptic3}
\end{align}
\end{subequations}
where we have used the identity $\cd(ka,-ik'/k)=\cn(a,k')$, and
\begin{align}
&\kappa\equiv \omega\sqrt{2H(I_y-I_x)}, k_1\equiv \frac{1}{k}= \sqrt{\frac{1-2HI_x}{2H(I_y-I_x)}}, \nonumber\\
&k'_1\equiv -\frac{ik'}{k}=\sqrt{\frac{2HI_y-1}{2H(I_y-I_x)}}.
\end{align}
Here, $\cd u$, $\nd u$ and $\sd u$ are auxiliary elliptic functions defined by $\cd u\equiv \cn u/\dn u$, $\nd u\equiv 1/\dn u$, and $\sd u\equiv \sn u/\dn u$. If Eq.\:\eqref{EulerTopFinalExpression1} -- Eq.\:\eqref{EulerTopFinalExpression3} and Eq.\:\eqref{FinalElliptic1} -- Eq.\:\eqref{FinalElliptic3} are transformed back into the original coordinate system, for $2HI_y\leq 1$ we have
\begin{subequations}\label{FinalClassicalSpinSmall}
\begin{align}\label{FinalClassicalSpinSmall1}
&S_x=\cos\theta\cn(a,k')\cn(\lambda t,k)+\sin\theta\dn(a,k')\sn(\lambda t,k),\\
&S_y=\cos\theta\dn(a,k')\sn(\lambda t,k)-\sin\theta\cn(a,k')\cn(\lambda t,k),\\
&S_z=\sn(a,k')\dn(\lambda t,k),\label{FinalClassicalSpinSmall3}
\end{align}
\end{subequations}
whereas for $2HI_y\geq 1$, we have
\begin{subequations}\label{FinalClassicalSpinLarge}
\begin{align}\label{FinalClassicalSpinLarge1}
S_x&=\cos\theta\cd(ka,k'_1)\dn(\kappa t,k_1)+k_1\sin\theta \nd(ka,k'_1)\sn(\kappa t,k_1),\\
S_y&=k_1\cos\theta \nd(ka,k'_1)\sn(\kappa t,k_1)-\cos\theta\cd(ka,k'_1)\dn(\kappa t,k_1),\\
S_z&=k_1\sd(ka,k'_1)\cn(\kappa t,k_1).\label{FinalClassicalSpinLarge3}
\end{align}
\end{subequations}
As a remark, the energies of the classical spin at the fixed points $M_x=\pm 1$ and $M_y=\pm 1$ are $H_{max}\equiv 1/(2I_x)$ and $H_{sep}\equiv 1/(2I_y)$ respectively, where $H_{max}$ denotes the maximum energy of the classical spin for fixed parameters, and $H_{sep}$ represents the energy for the separatrix which passes through the point $M_y=1$ and $M_y=-1$. In other words, under the condition $\alpha\gamma>\beta^2$ and $H<H_{max}$, Eq.\:\eqref{FinalClassicalSpinSmall1} -- Eq.\:\eqref{FinalClassicalSpinSmall3} and Eq.\:\eqref{FinalClassicalSpinLarge1} -- Eq.\:\eqref{FinalClassicalSpinLarge3} are responsible for $H\leq H_{sep}$ and $H\geq H_{sep}$ respectively. 
\begin{figure}[htbp]
\centering
\fbox{\includegraphics[width=0.98\linewidth]{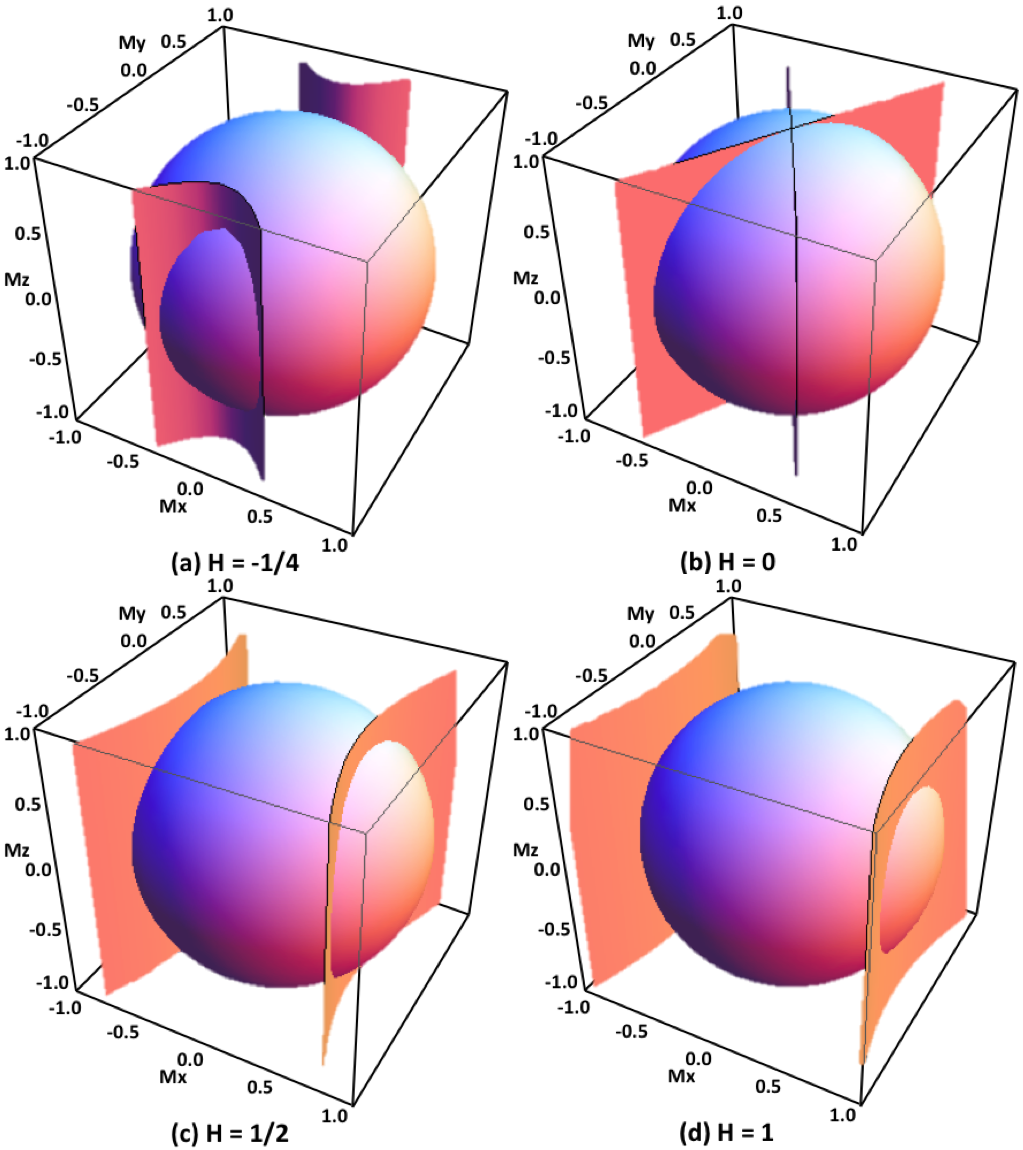}}
\caption{Intersection of the hyperbolic cylinder $M_x^2/J_x-M_y^2/J_y=2H$ with the unit sphere $M_x^2+M_y^2+M_z^2=1$ at different energies $H$. Here we set $\alpha=\gamma=1$ and $\beta=2$, so that $J_x=1/3$ and $J_y=2$. There are intersections between the hyperbolic cylinder and the unit sphere if and only if the condition $-1/2<H<3/2$ is fulfilled.}\label{fig:HyperbolicCylinderAndSphere}
\end{figure}
Specifically, for $H\approx 0$, the elliptic modulus $k\equiv \sqrt{2H(I_y-I_x)/(1-2HI_x)}\approx 0$, and we can approximate the Jacobi elliptic functions by circular functions: $\sn u\approx\sin u$, $\cn u\approx\cos u$, $\dn\tau\approx 1$. Hence, in the vicinity of the $S_z$ axis, we have
\begin{subequations}
\begin{align}
S_x&\approx\cos\theta\sqrt{2HI_x}\cos u+\sin\theta\sqrt{2HI_y}\sin u, \\
S_y&\approx-\sin\theta\sqrt{2HI_x}\cos u+\cos\theta\sqrt{2HI_y}\sin u.
\end{align}
\end{subequations}
Similarly, for $H\approx H_{sep}$, $k\equiv \sqrt{2H(I_y-I_x)/(1-2HI_x)}\approx 1$, and we can approximate the Jacobi elliptic functions by the hyperbolic functions: $\sn u\approx \tanh u$, $\cn u\approx \dn u \approx \sech u$. Hence, near the separatrix, we have
\begin{subequations}
\begin{align}
S_x&\approx\cos\theta\sqrt{I_x/I_y}\sech u+\sin\theta\tanh u,\\
S_y&\approx-\sin\theta\sqrt{I_x/I_y}\sech u+\cos\theta\tanh u,\\
S_z&\approx\sqrt{1-I_x/I_y}\sech u.
\end{align}
\end{subequations}
The period of oscillation is infinite on the separatrix which connects the fixed points $M_y=\pm 1$.

\subsection{The motion of the spin for $\alpha\gamma<\beta^2$}
For the case $\alpha\gamma<\beta^2$, as we can seen from Eq.\:\eqref{AsymmetricEulerTop1} and Eq.\:\eqref{AsymmetricEulerTop2}, one of the principal moments of inertia becomes negative, and hence, the system under study is an inverted asymmetric Euler top, whose dynamics is described by the Hamiltonian
\begin{equation}
H =  \frac{M_x^2}{2J_x}-\frac{M_y^2}{2J_y},
\end{equation}
where $J_x\equiv I_x$ and $J_y\equiv -I_y$ are two positive parameters. In terms of $M_x$, $M_y$ and $M_z$, the laws of conservation of energy and angular momentum become
\begin{subequations}
\begin{align} \label{HyperbolicCylinder1}
2H&=M_x^2/J_x-M_y^2/J_y,\\
1&=M_x^2+M_y^2+M_z^2.\label{HyperbolicCylinder2}
\end{align}
\end{subequations}
Eq.\:\eqref{HyperbolicCylinder1} and Eq.\:\eqref{HyperbolicCylinder2} are respectively the equations of a hyperbolic cylinder and that of a unit sphere. The existence of an intersection is ensured by the conditions $H_{min}<H<H_{max}$, where $H_{max}=(2J_x)^{-1}$ and $H_{min}=-(2J_y)^{-1}$. Fig.~\ref{fig:HyperbolicCylinderAndSphere} shows the lines of intersection of the hyperbolic cylinder with the unit sphere at various energies. Similar to the case of $\alpha\gamma>\beta^2$, the system contains six fixed points, which are respectively $M_x=\pm 1$, $M_y=\pm 1$, and $M_z=\pm 1$. However, the nature of the motion surrounding the various fixed points are different. The motion is stable near the $M_x$ and $M_y$-axes, but unstable near the $M_z$-axis. The hyperbolic saddles at $M_z=\pm 1$ are connected by four heteroclinic orbits with $H=0$, which satisfy the condition $M_x/M_y=\pm\sqrt{J_x/J_y}$. Using Eq.\:\eqref{EOM}, we obtain the equations of motion of the asymmetric top in terms of $M_x$, $M_y$ and $M_z$
\begin{equation}\label{EquationsOfMotionForHyperbolicTop}
\dot{M}_x=\frac{M_yM_z}{J_y},\dot{M}_y=\frac{M_xM_z}{J_x},\dot{M}_z=-\left(\frac{1}{J_x}+\frac{1}{J_y}\right)M_xM_y.
\end{equation}
According to Eq.\:\eqref{HyperbolicCylinder1} and Eq.\:\eqref{HyperbolicCylinder2}, we can express $M_x$ and $M_y$ in terms of $M_z$ as
\begin{subequations}
\begin{align}\label{Jx}
M_x^2 &= \frac{J_x}{J_x+J_y}[(1+2HJ_y)-M_z^2],\\
M_y^2 &= \frac{J_y}{J_x+J_y}[(1-2HJ_x)-M_z^2],\label{Jy}
\end{align}
\end{subequations}
where $1-2HJ_x\leq M_z^2\leq 1+2HJ_y$. Substitution of Eq.\:\eqref{Jx} and Eq.\:\eqref{Jy} into Eq.\:\eqref{EquationsOfMotionForHyperbolicTop} immediately yields
\begin{subequations}\label{HyperbolicEquationOfMotion}
\begin{align}\label{HyperbolicEquationOfMotion1}
\dot{M}_x &= \Omega\sqrt{(M_x^2-2HJ_x)[1+2HJ_y-(J_x+J_y)M_x^2/J_x)]},\\
\dot{M}_y &= \Omega\sqrt{(2HJ_y+M_y^2)[1-2HJ_x-(J_x+J_y)M_y^2/J_y]},\\
\dot{M}_z &= -\Omega\sqrt{[(1-2HJ_x)-M_z^2][(1+2HJ_y)-M_z^2]},\label{HyperbolicEquationOfMotion3}
\end{align}
\end{subequations}
where $\Omega\equiv \sqrt{\beta^2-\alpha\gamma}$. In the following, we discuss the two cases $H\leq 0$ and $H\geq 0$ separately. For the case $H\geq 0$, we introduce a set of new variables $u\equiv\Omega t\sqrt{1+2HJ_y}$, $m_x\equiv k'M_x/\sqrt{2HJ_x}$, $m_y\equiv k'M_y/(k\sqrt{2HJ_y})$, $m_z\equiv M_z/\sqrt{1-2HJ_x}$, $k^2\equiv (1-2HJ_x)/(1+2HJ_y)$, and $k'^2\equiv 2H(J_x+J_y)/(1+2HJ_y)$, so that Eq.\:\eqref{HyperbolicEquationOfMotion1} -- Eq.\:\eqref{HyperbolicEquationOfMotion3} become
\begin{subequations}\label{Hyperbolic}
\begin{align}
\frac{dm_x}{du} &= \sqrt{(1-m_x^2)(m_x^2-k'^2)},\\
\frac{dm_y}{du} &= \sqrt{(1-m_y^2)(k'^2+k^2m_y^2)},\\
\frac{dm_z}{du} &= -\sqrt{(1-m_z^2)(1-k^2m_z^2)}.
\end{align}
\end{subequations}
Hence, the components of $\mathbf{M}$ can be expressed in terms of the Jacobi elliptic functions as (we assume for definiteness that $M_x>0$)
\begin{subequations}\label{InvertedEulerTopFinalExpression}
\begin{align}
M_x&=\sqrt{\frac{J_x(1+2HJ_y)}{J_x+J_y}}\dn (\xi t,k)=\sn(\vartheta,k')\dn (\xi t,k),\\
M_y&=\sqrt{\frac{J_y(1-2HJ_x)}{J_x+J_y}}\cn (\xi t,k)=\cn(\vartheta,k')\cn (\xi t,k),\\
M_z&=-\sqrt{1-2HJ_x}\sn (\xi t,k)=-\dn(\vartheta,k')\sn (\xi t,k),
\end{align}
\end{subequations}
where $\xi\equiv \Omega\sqrt{1+2HJ_y}$, and $\vartheta$ is a real parameter defined by $\dn(\vartheta,k')\equiv \sqrt{1-2HJ_x}$, which satisfies $\sd(\vartheta,k')=\sin\vartheta/k$ and $\cd(\vartheta,k')=\cos\vartheta$.

For the case $H\leq 0$,, the elliptic modulus $k$ is larger than $1$, and we can again apply the identities for elliptic modulus outside the interval $[0,1]$: $\sn(u,k)=k^{-1}\sn(ku,k^{-1})$, $\cn(u,k)=\dn(ku,k^{-1})$ and $\dn(u,k)=\cn(ku,k^{-1})$, so that the components of $\mathbf{M}$ can be expressed in terms of the Jacobi elliptic functions as (we assume for definiteness that $M_y> 0$)
\begin{subequations}\label{InvertedEulerTopFinalExpression2}
\begin{align}
M_x&=\sqrt{\frac{J_x(1+2HJ_y)}{J_x+J_y}}\cn (\rho t,k_1)=\cn(\vartheta_1,k'_1)\cn (\rho t,k_1),\\
M_y&=\sqrt{\frac{J_y(1-2HJ_x)}{J_x+J_y}}\dn (\rho t,k_1)=\sn(\vartheta_1,k'_1)\dn (\rho t,k_1),\\
M_z&=-\sqrt{1+2HJ_y}\sn (\rho t,k_1)=\dn(\vartheta_1,k'_1)\sn (\rho t,k_1),
\end{align}
\end{subequations}
where $\rho\equiv \Omega \sqrt{1-2HJ_x}$, $k_1\equiv 1/k$, $k'_1\equiv\sqrt{1-k_1^2}$, and $\vartheta_1$ is a real number defined by $\dn(\vartheta_1,k'_1)\equiv \sqrt{1+2HI_y}$. In particular, for $H=0$ and $k=\tilde{k}=1$, the Jacobi elliptic functions degenerate into the hyperbolic functions: $\sn\tau\rightarrow\tanh\tau$, $\cn\tau\rightarrow\sech\tau$, $\dn\tau\rightarrow\sech\tau$. Hence, we obtain
\begin{subequations}
\begin{align}
M_x&=\sin\vartheta\sech \Omega t,\\
M_y&=\cos\vartheta\sech \Omega t,\\
M_z&=-\tanh \Omega t,
\end{align}
\end{subequations}
where $\sin\vartheta\equiv \sqrt{J_x/(J_x+J_y)}$. The above formulas give the explicit form of the heteroclinic orbits that connect the hyperbolic saddles at $M_z=\pm 1$. For $H \geq 0$, the components of classical spin $\mathbf{S}$ can be expressed as
\begin{subequations}\label{FinalClassicalSpinPositiveInverted}
\begin{align}
S_x&=\cos\theta\sn(\vartheta,k')\dn (\xi t,k)+\sin\theta\cn(\vartheta,k')\cn (\xi t,k),\\
S_y&=-\sin\theta\sn(\vartheta,k')\dn (\xi t,k)+\cos\theta\cn(\vartheta,k')\cn (\xi t,k),\\
S_z&=-\dn(\vartheta,k')\sn (\xi t,k),
\end{align}
\end{subequations}
whereas for $H\leq 0$, we obtain
\begin{subequations}\label{FinalClassicalSpinNegativeInverted}
\begin{align}
S_x&=\cos\theta\cn(\vartheta_1,k'_1)\cn (\rho t,k_1)+\sin\theta\sn(\vartheta_1,k'_1)\dn (\rho t,k_1),\\
S_y&=-\sin\theta\cn(\vartheta_1,k'_1)\cn (\rho t,k_1)+\cos\theta\sn(\vartheta_1,k'_1)\dn (\rho t,k_1),\\
S_z&=-\dn(\vartheta_1,k'_1)\sn (\rho t,k_1).
\end{align}
\end{subequations}
In particular, for $H=0$, we obtain
\begin{subequations}
\begin{align}
S_x&=\sin(\vartheta+\theta)\sech \Omega t,\\
S_y&=\cos(\vartheta+\theta)\sech \Omega t,\\
S_z&=-\tanh \Omega t.
\end{align}
\end{subequations}
The area enclosed by the heteroclinic orbits passing through the hyperbolic saddles at $S_z=\pm 1$ may be calculated from elementary geometry. For definiteness, we discuss the two heteroclinic orbits with $M_y>0$. Let us denote the heteroclinic orbit with $M_x>0$ as $l_1$, and the heteroclinic orbit with $M_x<0$ as $l_2$. The angle between $l_1$ and the $M_x$-axis is given by $\tan\theta_1=\sqrt{J_x/J_y}$. Similarly, the angle between $l_1$ and the $M_x$-axis is given by $\tan\theta_1=-\sqrt{J_x/J_y}$. Hence, the angle between $l_1$ and $l_2$ is given by
\begin{equation}
\theta_2-\theta_1=\pi-2\arctan(\sqrt{J_x/J_y}),
\end{equation}
which implies that the area enclosed by $l_1$ and $l_2$ is $2\pi-4\arctan(\sqrt{J_x/J_y})$. In particular, this area becomes $\pi$ when $J_x=J_y$.

\section{Conclusion and Discussion}
In this work, we explore the nonlinear fiber-optic correspondence of excited-state quantum phase transitions within the generalized Lipkin-Meshkov-Glick (LMG) model, across different fiber media and symmetry classes. Beyond the conventional LMG model, which includes a rotor-like term and a single-axis quadratic squeezing term, we identify an additional term in the nonlinear polarization dynamics for tetragonal symmetry classes. This term corresponds to an unconventional squeezing mechanism within the generalized LMG framework, offering new insights into the interplay between classical and quantum dynamics in structured optical systems.

Our analysis provides a detailed discussion of classical bifurcation phenomena in nonlinear polarization dynamics. A key finding is that, even in the absence of the rotor-like term, a bifurcation occurs where the energy landscape—expressed as a function of spin components—transitions from an ellipsoidal shape to a hyperbolic one. This transition may signify a new type of quantum phase transition in its quantum counterpart and could lead to novel squeezed states beyond conventional forms. These extended investigations, particularly the exploration of these novel squeezed states, will be presented in a separate paper.

We focus on the nonlinear polarization dynamics of two birefringent modes in a nonlinear optical fiber composed of different materials with varying symmetry classes. While our scope is well-defined, one could extend the analysis to a more general case where higher-order nonlinear susceptibilities play a role. In such a scenario, the polarization dynamics on the Poincar\'e sphere would be governed by a more complex Hamiltonian, incorporating higher-order products of spin generators. This leads to the formulation of the following mathematical problem:
\begin{subequations}
\begin{align}
\dot{S}_i &= \{S_i, H\}, \quad i \in \{x, y, z\}, \\
H &= \sum_{n=1}^N \sum_{i+j+k \leq n} a_{ijk}^{(n)} S_x^i S_y^j S_z^k, \\
1 &= S_x^2 + S_y^2 + S_z^2,
\end{align}
\end{subequations}
where $H$ is the generalized Hamiltonian, and the summations include terms up to order $N$, accounting for the nonlinear interactions among spin components $S_x$, $S_y$, and $S_z$, and the parameters $a_{ijk}^{(n)}$ represent material-dependent parameters and symmetry constraints. The structural transition in classical spin dynamics on the Poincar\'e sphere is characterized by the appearance or disappearance of equilibrium points, as described by bifurcation theory. These sudden structural changes are determined by a set of discriminants of the parameters $a_{ijk}^{(n)}$, an approach rooted in algebraic geometry and classical invariant theory \cite{olver1999classical}. This perspective has also been utilized in quantum computation to classify entanglement \cite{kam2022genuine}, highlighting the interdisciplinary relevance of our findings.

A compelling theoretical question arises: to what extent can these emergent bifurcation behaviors be mapped onto excited-state quantum phase transitions? The latter involves abrupt changes in spectral structures and entanglement properties. One possible direction to explore would be how the discriminants governing bifurcations relate to quantum order parameters, such as entanglement entropy or fidelity susceptibility. If a rigorous mapping can be established, it could provide new insights into how symmetry-breaking phenomena in classical systems mirror critical behavior in quantum mechanics, potentially unifying classical and quantum perspectives on critical phenomena.

Experimentally, a key question is how—and to what extent—the most generalized case can be realized. From the nonlinear optical analogs of the generalized Lipkin-Meshkov-Glick (LMG) model, we found that not all symmetry classes of fiber materials correspond to conserved Hamiltonian dynamics. Typically, additional constraints must be satisfied. Furthermore, along the fiber direction, the system supports at most two birefringent modes, limiting the experimental configuration. Alternative approaches, such as coupled fiber systems, may offer a more comprehensive way to explore the full complexity of the problem, potentially enabling the simulation of multipartite quantum phenomena like entanglement or topological order.

On the quantum side, promising platforms for investigating these effects include spinor Bose-Einstein condensates (BECs) and trapped ion systems, which provide controllable environments to study the interplay of nonlinear and quantum dynamics. These platforms could validate the optical analogues by offering precise control over quantum interactions, complementing the room-temperature accessibility of optical fibers. Collectively, our findings position nonlinear optical systems as a versatile bridge between classical and quantum physics, with transformative potential for quantum metrology, simulation, and the exploration of non-equilibrium phenomena like time crystals.

\begin{appendix}
\section{Scalar nonlinear Schr\"{o}dinger equation (NLSE) in a nonlinear optical fibers}\label{A}
Like all electromagnetic phenomena in nature, the propagation of electromagnetic fields in a dispersive nonlinear dielectric medium is governed by Maxwell's equations \cite{agrawal2000nonlinear}
\begin{equation}\label{MaxwellEquationTime}
\nabla\times\nabla\times\mathbf{E}(\mathbf{r},t)+\frac{1}{\epsilon_0c^2}\frac{\partial^2\mathbf{D}(\mathbf{r},t)}{\partial t^2}=0,
\end{equation}
where $\epsilon_0$ is the free-space permittivity, $\mathbf{D}(\mathbf{r},t)=\epsilon_0\mathbf{E}(\mathbf{r},t)+\mathbf{P}(\mathbf{r},t)$ is the electric displacement, and $\mathbf{P}(\mathbf{r},t)$ is the induced electric polarization of the medium in response to the applied electromagnetic field.  As fiber optics are usually made of nonmagnetic media, the induced magnetic polarization $\mathbf{M}(\mathbf{r},t)$ may be neglected.  For sufficiently weak electric fields, the induced electric polarization $\mathbf{P}(\mathbf{r},t)$ can be expanded into a power series of the applied electric field \cite{shen1984principles}
\begin{subequations}
\begin{gather}
\mathbf{P}(\mathbf{r},t)=\mathbf{P}_L(\mathbf{r},t)+\mathbf{P}_{NL}(\mathbf{r},t),\\
\mathbf{P}_L(\mathbf{r},t)=\epsilon_0\int_Vd\mathbf{r}_1\int_{-\infty}^t dt_1\chi^{(1)}(\mathbf{r}-\mathbf{r}_1,t-t_1)\cdot\mathbf{E}(\mathbf{r}_1,t_1),\\
\mathbf{P}_{NL}(\mathbf{r},t)=\epsilon_0\int_Vd\mathbf{r}_1\int_Vd\mathbf{r}_2\int_Vd\mathbf{r}_3\int_{-\infty}^t dt_1 \int_{-\infty}^t dt_2 \int_{-\infty}^t dt_3 \nonumber\\ 
\chi^{(3)}(\mathbf{r}-\mathbf{r}_1,t-t_1,\mathbf{r}-\mathbf{r}_2,t-t_2,\mathbf{r}-\mathbf{r}_3,t-t_3)\nonumber\\ 
\:\vdots\:\mathbf{E}(\mathbf{r}_1,t_1)\mathbf{E}(\mathbf{r}_2,t_2)\mathbf{E}(\mathbf{r}_3,t_3),
\end{gather}
\end{subequations}
where $\mathbf{P}_L(\mathbf{r},t)$ and $\mathbf{P}_{NL}(\mathbf{r},t)$ are the linear and nonlinear parts of the induced electric polarization respectively, $\chi^{(1)}(\mathbf{r}-\mathbf{r}_1,t-t_1)$ is the linear response function that describes the linear part of the induced polarization at point $\mathbf{r}$ and time $t$ in response to the applied electric fields at point $\mathbf{r}_1$ and time $t_1$, and $\chi^{(3)}(\mathbf{r}-\mathbf{r}_1,t-t_1,\mathbf{r}-\mathbf{r}_2,t-t_2,\mathbf{r}-\mathbf{r}_3,t-t_3)$ is the third order response function that describes the response of the medium at point $\mathbf{r}$ and time $t$ resulting from the applied electric field at point $\mathbf{r}_1$ and time $t_1$, point $\mathbf{r}_2$ and time $t_2$, and point $\mathbf{r}_3$ and time $t_3$.  Here, the tensor notation is defined as $(\chi^{(3)}\:\vdots\:\mathbf{E}\mathbf{E}\mathbf{E})_i\equiv \sum_{j,k,l}\chi^{(3)}_{ijkl}E_jE_kE_l$.  In general, the effect of spatial dispersion in normal crystals is weak, since the linear response function $\chi^{(1)}(\mathbf{r}-\mathbf{r}_1,t-t_1)$ reaches its maximum value at $\mathbf{r}=\mathbf{r}_1$ and drops quickly for $|\mathbf{r}-\mathbf{r}_1|>a$ for some characteristic dimension, $a$, of the medium.  The third order response function $\chi^{(3)}(\mathbf{r}-\mathbf{r}_1,t-t_1,\mathbf{r}-\mathbf{r}_2,t-t_2,\mathbf{r}-\mathbf{r}_3,t-t_3)$ experiences similar localization effects.  Here, the strength of spatial dispersion is measured by the ratio $a/\lambda$, where $\lambda$ is the wavelength of the optical wave in the medium.  For normal crystals, $a$ is on the order of the lattice constant, that is, $a\approx 1$nm, and thus the parameter $a/\lambda$ is a small quantity even for $\lambda\approx 1\mu$m, where $a/\lambda\approx 10^{-3}$ \cite{agranovich1984crystal}.  Hence, we neglect spatial dispersion effects, and only consider the local response of the medium, so that the electric polarization in response to the applied electric field can be approximated by
\begin{subequations}\label{NonlinearPolarizationWithoutSpatialEffects}
\begin{gather}\label{NonlinearPolarizationWithoutSpatialEffects1}
\mathbf{P}_L(\mathbf{r},t)\approx\epsilon_0\int_{-\infty}^t dt_1\chi^{(1)}(t-t_1)\cdot\mathbf{E}(\mathbf{r},t_1),\\
\mathbf{P}_{NL}(\mathbf{r},t)\approx\epsilon_0\int_{-\infty}^t dt_1 \int_{-\infty}^t dt_2 \int_{-\infty}^t dt_3 \nonumber\\ 
\chi^{(3)}(t-t_1,t-t_2,t-t_3)\:\vdots\:\mathbf{E}(\mathbf{r},t_1)\mathbf{E}(\mathbf{r},t_2)\mathbf{E}(\mathbf{r},t_3),\label{NonlinearPolarizationWithoutSpatialEffects2}
\end{gather}
\end{subequations}
where $\chi^{(1)}(t-t_1)$ is the linear susceptibility and $\chi^{(3)}(t-t_1,t-t_2,t-t_3)$ is the third-order nonlinear susceptibility. 

In the absence of nonlinearity, Maxwell's equations \eqref{MaxwellEquationTime} can be solved by Fourier transform, which yields \cite{snyder1983single}
\begin{equation}\label{MaxwellEquation}
\nabla\times\nabla\times\tilde{\mathbf{E}}(\mathbf{r},\omega)-\epsilon(\omega)\frac{\omega^2}{c^2}\tilde{\mathbf{E}}(\mathbf{r},\omega)=0,
\end{equation}
where $\epsilon(\omega)\equiv 1+\tilde{\chi}^{(1)}(\omega)$ is the frequency-dependent dielectric function and $\tilde{\mathbf{E}}(\mathbf{r},\omega)$ is the Fourier transform of $\mathbf{E}(\mathbf{r},t)$ defined by
\begin{equation}
\tilde{\mathbf{E}}(\mathbf{r},\omega) \equiv \int_{-\infty}^{\infty} \mathbf{E}(\mathbf{r},t)e^{i\omega t} dt.
\end{equation}
Here, $\tilde{\chi}^{(1)}(\omega)$ is the Fourier transform of $\chi^{(1)}(t-t_1)$ defined by
\begin{equation}
\tilde{\chi}^{(1)}(\omega) \equiv \int_{t_1}^\infty  \chi^{(1)}(t-t_1)e^{i\omega(t-t_1)}dt =\int_0^{\infty} \chi^{(1)}(\tau) e^{i\omega \tau} d\tau.
\end{equation}
In accordance with the local susceptibility approximation, we may write 
\begin{equation}
\nabla\times\nabla\times\tilde{\mathbf{E}}\equiv \nabla(\nabla\cdot\tilde{\mathbf{E}})-\nabla^2\tilde{\mathbf{E}} \approx -\nabla^2\tilde{\mathbf{E}},
\end{equation}
where $\nabla^2\tilde{\mathbf{E}}$ is the vector Laplacian acting on the electric field $\tilde{\mathbf{E}}$, and we have used Gauss's law $\nabla\cdot\tilde{\mathbf{E}} \approx \epsilon^{-1}\nabla\cdot\tilde{\mathbf{D}} =  0$ in the last step. Hence, the electric field in a linear optical fiber in the frequency domain, $\tilde{\mathbf{E}}(\mathbf{r},\omega)$, is governed by the Helmholtz equation \cite{agrawal2000nonlinear}
\begin{equation}\label{HelmholtzEquation}
\nabla^2\tilde{\mathbf{E}}(\mathbf{r},\omega)+\epsilon(\omega)k^2\tilde{\mathbf{E}}(\mathbf{r},\omega)=0,
\end{equation}
where $k\equiv \omega/c$.  Because of the cylindrical symmetry of the fibers, we may express the wave equation \eqref{HelmholtzEquation} in cylindrical coordinates $(r,\phi,z)$, and the $z$-component of the electric field is found to obey
\begin{equation}\label{ZHelmholtzEquation}
\left[\frac{\partial^2}{\partial r^2}+\frac{1}{r}\frac{\partial}{\partial r}+\frac{1}{r^2}\frac{\partial^2}{\partial \phi^2}+\frac{\partial^2}{\partial z^2}+\epsilon(\omega)k_0^2\right]\tilde{E}_z(r,\phi,z,\omega)=0.
\end{equation}
As a remark, there are six components of the electromagnetic field in a dielectric medium --- $\tilde{E}_x$, $\tilde{E}_y$, $\tilde{E}_z$, $\tilde{H}_x$, $\tilde{H}_y$ and $\tilde{H}_z$. However, only $\tilde{E}_z$ and $\tilde{H}_z$ are independent components, as all other components of the electromagnetic field can be calculated from Maxwell's equations.  The Helmholtz equation \eqref{ZHelmholtzEquation} can be solved by the standard method of separation of variables by writing $\tilde{E}_z(r,\phi,z,\omega)=A(\omega)F(r)e^{i(m\phi+\beta z)}$, where $m$ is an integer. A direct computation yields
\begin{equation}\label{HelmholtzEquationRadical}
\left[\frac{d^2}{dr^2}+\frac{1}{r}\frac{d}{dr}-\frac{m^2}{r^2}-\beta^2+\epsilon(\omega)k_0^2\right]F(r) = 0.
\end{equation}
For a given frequency $\omega$ of the optical field, the solutions of the Helmholtz equation depend on the fiber characteristics.  An optical fiber contains a core and a cladding layer, where the core has a refractive index $n_1$ slightly higher than the refractive index of the cladding $n_c$.  In general, the fiber core is made by an extremely low loss medium, such as silica glass, and hence fiber loss can be neglected in calculating the fiber modes. Thus, the frequency-dependent dielectric function $\epsilon(\omega)$ can be approximated by $n_1^2(\omega)$ inside the fiber core, and by $n_c^2(\omega)$ in the cladding.  As a result, the Helmholtz equation \eqref{HelmholtzEquationRadical} for $F(r)$ takes the form
\begin{subequations}\label{BesselFunction}
\begin{gather}
r^2F''(r)+rF'(r)+(p^2r^2-m^2)F(r) = 0,\:(r\leq a); \label{Bessel1}\\
r^2F''(r)+rF'(r)-(q^2r^2+m^2)F(r) = 0,\:(r\geq a),\label{Bessel2}
\end{gather}
\end{subequations}
where $p^2\equiv n_1^2k^2-\beta^2$ and $q^2\equiv \beta^2-n_c^2k^2$. \eqref{Bessel1} and \eqref{Bessel2} are solved by the Bessel functions $J_m(pr)$ and $K_m(qr)$: $F(r)=J_m(pr)$ for $r\leq a$, and $F(r)=K_m(pr)$ for $r\geq a$. Here, $J_m(u)$ is the Bessel function of the first kind defined by the integral representation \cite{courant1965methods}
\begin{equation}
J_m(u) \equiv \frac{1}{2\pi}\int_{-\pi}^{\pi}\mathcal{J}(u,\tau)e^{im\tau}d\tau,
\end{equation}
where $\mathcal{J}(u,\tau)\equiv e^{-iu\sin\tau}$ is the integration kernel, and $K_m(u)$ is the modified Bessel function of the second kind defined by integral representation \cite{courant1965methods}
\begin{equation}
K_m(u)\equiv \int_0^\infty \mathcal{K}(u,\tau)e^{-m\tau}d\tau,
\end{equation}
where $\mathcal{K}(u,\tau)\equiv e^{-u\sinh\tau}$ is the integration kernel.  As remark, a general solution of \eqref{Bessel1} is a linear combination of $J_m(pr)$ and $N_m(pr)$, where $N_m(u)$ is the Bessel function of the second kind.  However, as $N_m(u)$ presents a singularity at $u=0$, it is not appropriate for a solution inside the core.  Similarly, a general solution of \eqref{Bessel2} contains a linear combination of $I_m(u)$ and $K_m(u)$, where $I_m(u)$ is the modified Bessel function of the first kind.  Here, only $K_m(u)$ decays exponentially at large values of $u$, which is appropriate for a solution in the cladding region.  Similar procedures can be applied to the $z$-component of the magnetic field $\tilde{H}_z$.  Finally, the propagation constant $\beta(\omega)$ in \eqref{HelmholtzEquationRadical} is determined by the boundary conditions for the electromagnetic fields --- the tangential components of $\tilde{\mathbf{E}}$ and $\tilde{\mathbf{H}}$ should be continuous across the core-cladding boundary, so that $\tilde{E}_z$, $\tilde{E}_\phi$, $\tilde{H}_z$ and $\tilde{H}_\phi$ have the same values on both sides of the boundary $r=a$.  For each given value of $m$, the solution for $\beta$ is not unique, and thus $\beta$ is in general labeled by an integer $n$.  Hence, for a given frequency $\omega$, there exist a series of discrete modes of the electromagnetic field inside the fiber, and each specific mode supported by the fiber is characterized by a propagation constant $\beta_{mn}$.  Unlike the transverse-electric (TE) and transverse-magnetic (TM) modes in a planar waveguide, for $m>0$, the modal fields supported by an optical fiber are in general hybrid, that is, both $E_z$ and $H_z$ are nonzero, and hence these modes are denoted by HE$_{mn}$ or EH$_{mn}$ depending on which components of $\tilde{\mathbf{E}}$ or $\tilde{\mathbf{H}}$ have greater contribution to the transverse field.  Nevertheless, the analysis of modal fields can be simplified in telecommunication-grade optical fibers, since the weakly guiding approximation is valid, that is, the difference between the refractive indices of the core and cladding is much smaller than the refractive index of the core, i.e., $(n_1-n_c)/n_1<0.01$, so that the propagation of the electromagnetic field is essentially along the fiber axis, which yields predominantly TE and TM modes.  Thus, the fiber modes are approximated by two linearly polarized components in the transverse directions, as $E_z\approx H_z\approx 0$, and the linearly polarized modes are denoted by LP$_{mn}$.  Let the optical field be linearly polarized in the $x$-direction, then the fundamental mode LP$_{01}$ can be explicitly written as \cite{gloge1971weakly, agrawal2000nonlinear}
\begin{subequations}\label{TransverseDistribution}
\begin{align}\label{TransverseDistribution1}
\tilde{E}_x(r,z,\omega)&= A(\omega)J_0(pr)/J_0(pa)e^{i\beta(\omega)z}\:\:\;\;\mbox{for}\;\:r\leq a;\\
\tilde{E}_x(r,z,\omega)&= A(\omega)K_0(qr)/K_0(qa)e^{i\beta(\omega)z}\:\;\mbox{for}\:\;r\geq a.\label{TransverseDistribution2}
\end{align}
\end{subequations}
The number of modes inside an optical fiber at a specific frequency depends on the core radius $a$ and the difference between the refractive indices of the core and the cladding, that is, $n_1-n_c$.  The cut-off frequency for a given linear polarized mode LP$_{mn}$ is determined by the cut-off condition $J_{m-1}(V)=0$, where $V\equiv ka\sqrt{n_1^2-n_c^2}$.  Specifically, for $V<V_c\approx 2.405$, all higher fiber modes disappear, except for the fundamental mode.  For example, if $n_1-n_c = 0.005$ and $a=4\mu$m, the fiber supports a single mode LP$_{01}$ only when $\lambda>1.2\mu$m \cite{agrawal2000nonlinear}.  We have now discussed the linear polarized guided modes LP$_{mn}$, which are supported by an optical fiber in the absence of nonlinearity. In particular, we have analyzed the fundamental mode LP$_{01}$ in detail.  In the following, we will discuss the pulse-propagation equation for a nonlinear optical fiber, where the weakly guiding approximation and the single-mode condition are assumed to be valid.

Besides these two main assumptions regarding the fiber characteristics, several additional assumptions should be made to ensure the existence of a nonlinear pulse-propagation equation.  First, the electric field is assumed to keep its polarization along the entire fiber length, as the two orthogonally polarized modes are degenerate in an ideal single-mode fiber.  Second, the optical field is assumed to be quasi-monochromatic, that is, we may express the electric field as the product of a slowly varying function in time and a fast oscillating sinusoid by writing
\begin{equation}\label{IsotropicElectricField}
\mathbf{E}(\mathbf{r},t) \equiv \mbox{Re}\{E(\mathbf{r},t)e^{-i\omega_0t}\}\hat{x},
\end{equation}
where the electric field is assumed to be polarized in the $x$-direction, $\omega_0\approx 10^{15}$\:s$^{-1}$ is the frequency of the fast oscillating part of the optical field, and $E(\mathbf{r},t)$ is a function varies slowly in time compared to $e^{-i\omega_0 t}$, which satisfies the condition $|\partial_{tt}E(\mathbf{r},t)|\ll |\omega_0\partial_tE(\mathbf{r},t)|$ \cite{shen1984principles}. Similarly, we can express the electric polarization as the product of a slowly varying function in time and a fast oscillating sinusoid by writing
\begin{subequations}\label{IsotropicPolarizations}
\begin{align}\label{IsotropicPolarizations1}
\mathbf{P}_L(\mathbf{r},t) &\equiv \mbox{Re}\{P_L(\mathbf{r},t)e^{-i\omega_0t}\}\hat{x}\:\;\mbox{and}\\
\mathbf{P}_{NL}(\mathbf{r},t) &\equiv \mbox{Re}\{P_{NL}(\mathbf{r},t)e^{-i\omega_0t}\}\hat{x}.\label{IsotropicPolarizations2}
\end{align}
\end{subequations}
Now, if we substitute \eqref{IsotropicElectricField} and \eqref{IsotropicPolarizations1} -- \eqref{IsotropicPolarizations2} into \eqref{MaxwellEquationTime} and \eqref{NonlinearPolarizationWithoutSpatialEffects1} -- \eqref{NonlinearPolarizationWithoutSpatialEffects2}, we can, in principal, obtain a propagation equation for the electric field.  However, such an analysis is in general complicated.  In order to obtain a simpler physical picture, we adopt the following approximation --- the induced nonlinear polarization in the medium in response to the applied electric field is assumed to be local in time, that is, the third-order nonlinear susceptibility takes the form of a product of three delta functions as
\begin{equation}\label{instantaneous}
\chi^{(3)}(t-t_1,t-t_2,t-t_3) \approx \chi^{(3)}\delta(t-t_1)\delta(t-t_2)\delta(t-t_3),
\end{equation}
where $\delta(t-t_i)$ are Dirac's delta functions.  In general, a non-instantaneous response can be caused by vibrations of molecules in the medium.  As the vibrations of electrons are much faster than those of nuclei, which typically involve a timescales of less than 10 fs, we can assume the electronic response to be nearly instantaneous.  On the other hand, the decay of nuclear response has a relaxation time ranging from 100 fs to 10 ps \cite{aber2000femtosecond}, and for silica, the delayed nuclear response usually occurs over a time scale 60 fs to 70 fs \cite{agrawal2000nonlinear}.  Hence, for ultrashort optical pulses whose durations are shorter than the relaxation times of the nuclei, the assumption of instantaneous response is no longer valid.  However, for optical pulses with durations lager than 5 ps, the vibrations of the molecules can be neglected, and both the electronic and nuclear responses can be regarded as instantaneous.  Substituting \eqref{instantaneous} into \eqref{NonlinearPolarizationWithoutSpatialEffects1} -- \eqref{NonlinearPolarizationWithoutSpatialEffects2}, we obtain the electric polarization regardless of molecular vibrations 
\begin{subequations}\label{ElectricPolarizationWithoutMolecularVibrations}
\begin{gather}
\mathbf{P}_L(\mathbf{r},t)=\epsilon_0\int_{-\infty}^t dt_1\chi^{(1)}(t-t_1)\cdot\mathbf{E}(\mathbf{r},t_1),\\
\mathbf{P}_{NL}(\mathbf{r},t)\approx\epsilon_0 \chi^{(3)}\:\vdots\:\mathbf{E}(\mathbf{r},t)\mathbf{E}(\mathbf{r},t)\mathbf{E}(\mathbf{r},t).\label{ThridOrderElectricPolarization}
\end{gather}
\end{subequations}
Dispersion effects arise whenever the linear response of a field is not instantaneous. Therefore, in order to take into account fiber dispersion effects, the linear response to the electric field is not taken as instantaneous. Substitution of \eqref{IsotropicElectricField} and \eqref{IsotropicPolarizations2} into \eqref{ThridOrderElectricPolarization} immediately yields
\begin{align}\label{ScalarNonlinearPolarization}
\mathbf{P}_{NL}(\mathbf{r},t) &= \frac{\epsilon_0}{4}\chi^{(3)}_{xxxx}\mbox{Re}\left\{E^3(\mathbf{r},t)e^{-3i\omega_0t}\right.\nonumber\\
&+\left.3|E(\mathbf{r},t)|^2E(\mathbf{r},t)e^{-i\omega_0t}\right\}\hat{x},
\end{align} 
where the term proportional to $e^{-3i\omega_0t}$ is associated with the nonlinear process of frequency tripling.  For example, in principle, if we input infrared 1024 nm light into a silica fiber, we can generate an ultraviolet 355 nm light through the nonlinear frequency tripling process.  However, nonlinear phenomena such as frequency doubling and frequency tripling require phase-matching techniques, that is, a specific fine-tuning of input wavelengths and fiber parameters.  Hence, the term proportional to $e^{-3i\omega_0t}$ is negligible in most cases, and we obtain
\begin{equation}\label{WithoutMatching}
P_{NL}(\mathbf{r},t) \approx \frac{3\epsilon_0}{4}\chi^{(3)}_{xxxx}|E(\mathbf{r},t)|^2E(\mathbf{r},t).
\end{equation}
Substituting \eqref{ElectricPolarizationWithoutMolecularVibrations} -- \eqref{ThridOrderElectricPolarization} and \eqref{WithoutMatching} into Maxwells's equations \eqref{MaxwellEquationTime}, we obtain a scalar wave equation for $E(\mathbf{r},t)$
\begin{align}\label{ScalarWaveEquation}
\nabla^2 E(\mathbf{r},t) e^{-i\omega_0t}&= \frac{1}{c^2}\frac{\partial^2}{\partial t^2}\left\{E(\mathbf{r},t)e^{-i\omega_0t} \right.\nonumber\\
&+\left.\int_{-\infty}^t dt_1\chi^{(1)}_{xx}(t-t_1) E(\mathbf{r},t_1)e^{-i\omega_0t_1}\right\}\nonumber\\
&+\frac{3\chi^{(3)}_{xxxx}}{4c^2}\frac{\partial^2}{\partial t^2}\left\{|E(\mathbf{r},t)|^2E(\mathbf{r},t)e^{-i\omega_0t}\right\},
\end{align}
where we have used the approximation $\nabla\cdot\mathbf{E} \approx \epsilon^{-1}\nabla\cdot\mathbf{D} =  0$ for a spatially uniform medium.  While \eqref{ScalarWaveEquation} provides a complete description of light propagation in optical fibers, it is still too complicated to be solved analytically.  In the following, we analyze the wave equation \eqref{ScalarWaveEquation} in the frequency domain. Here, the Fourier transform $\tilde{E}(\mathbf{r},\omega-\omega_0)$ of $E(\mathbf{r},t)$ is found to obey
\begin{equation}\label{NonHolmEq}
\nabla^2 \tilde{E}(\mathbf{r},\omega-\omega_0) + (\epsilon(\omega)+\delta\epsilon)k^2  \tilde{E}(\mathbf{r},\omega-\omega_0) =0,
\end{equation}
where $k\equiv \omega/c$ and $\epsilon(\omega)\equiv 1+\tilde{\chi}^{(1)}_{xx}(\omega)$. $\tilde{E}(\mathbf{r},\omega-\omega_0)$ is defined by
\begin{equation}
\tilde{E}(\mathbf{r},\omega-\omega_0) \equiv \int_{-\infty}^{\infty} E(\mathbf{r},t) e^{i(\omega-\omega_0)t}dt,
\end{equation}
and $\delta\epsilon\equiv \frac{3}{4}\chi^{(3)}_{xxxx}|E(\mathbf{r},t)|^2$ is the nonlinear contribution to the dielectric function.  In the derivation of \eqref{NonHolmEq}, we regarded $|E(\mathbf{r},t)|^2$ as a constant, so that the following approximation is used
\begin{equation}\label{NonlinearFouTransform}
\int_{-\infty}^{\infty} |E(\mathbf{r},t)|^2E(\mathbf{r},t)e^{i(\omega-\omega_0)t}dt \approx |E(\mathbf{r},t)|^2\tilde{E}(\mathbf{r},\omega-\omega_0).
\end{equation}
In general, the approximation \eqref{NonlinearFouTransform} is not rigorous.  Since Fourier transformation of a nonlinear function is mathematically ill-defined in most cases, it is meaningless to ask what the corresponding signals in the frequency domain are.  Moreover, the nonlinear part of the dielectric function, $\delta\epsilon$, varies with time, which leads to some ambiguities, as $\epsilon$ is generally considered to be a function of frequency \cite{haus1984waves}.  However, if we restrict our attention to the slowly varying function $E(\mathbf{r},t)$, and treat $\delta\epsilon$ as a small perturbation of $\epsilon(\omega)$, the approximation \eqref{NonlinearFouTransform} can be justified.  As a remark, the equations that govern light propagation in optical fibers can be derived rigorously using the reductive perturbation method \cite{hasegawa2003optical, taniuti1959wave, taniuti1974reductive}, and the result is consistent with the usage of nonlinear wave equation in the frequency domain.

In the absence of nonlinearity, as we have discussed in \eqref{Bessel1} -- \eqref{Bessel2} and their solutions therein, an optical fiber made of a linear medium supports a series of discrete guided modes, namely EH$_{mn}$ or HE$_{mn}$. In particular, under the weakly guiding condition, the refractive indices of the core and cladding are nearly the same, and hence the fiber modes are approximated by linearly polarized  LP$_{mn}$ modes, each of which comprises two polarizations \cite{snyder1983love, snyder1972coupled}.  A single-mode fiber only supports the fundamental LP$_{01}$ mode 
\begin{equation}
\tilde{E}(x,y,z,\omega) = \tilde{A}(\omega)F(x,y)e^{i\beta(\omega)z},
\end{equation}
where $\tilde{A}(\omega)$ is a normalization constant that depends only on the frequency and $\beta(\omega)$ is a propagation constant determined by
\begin{equation}\label{ModalDistribution}
[\nabla_t^2+n^2k^2-\beta^2]F=0,
\end{equation}
where $\nabla_t^2\equiv \partial_{xx}+\partial_{yy}$ is the transverse Laplacian operator and $n(\omega)$ is the refractive index of the linear medium.  Now, as we treat $\delta\epsilon\equiv \frac{3}{4}\chi^{(3)}_{xxxx}|E(\mathbf{r},t)|^2$ as a small perturbation of the dielectric function $\epsilon(\omega)$, we may solve \eqref{NonlinearFouTransform} by separation of variables, so that the slowly varying wave envelope of the electric field can be expressed in the form
\begin{equation}\label{ModalExpansion}
\tilde{E}(x,y,z,\omega-\omega_0) = \tilde{A}(z,\omega-\omega_0)F(x,y)e^{i\beta(\omega_0)z},
\end{equation}
where $\tilde{A}(z,\omega-\omega_0)$ is a slowly varying function of $z$ which will be determined later.  Substituting \eqref{ModalExpansion} into \eqref{NonHolmEq}, we obtain
\begin{subequations}\label{WaveEquation}
\begin{align}
\nabla_t^2F+[(\epsilon+\delta\epsilon)k^2-(\beta+\delta\beta)^2]F&=0,\label{NewModalDistribution} \\
2i\beta_0\frac{\partial \tilde{A}}{\partial z}+[(\beta+\delta\beta)^2-\beta_0^2]\tilde{A}&=0,\label{ModalAmplitude}
\end{align}
\end{subequations}
where $\beta_0\equiv \beta(\omega_0)$ and the second-order spatial derivative of $\tilde{A}(z,\omega-\omega_0)$ has been neglected in accordance with the slowly varying amplitude approximation \cite{snyder1978modes}.  Substitution of \eqref{ModalDistribution} into \eqref{NewModalDistribution} immediately yields
\begin{equation}\label{Perturbation}
[(\delta\epsilon+i\mbox{Im}\tilde{\chi}^{(1)}_{xx})k^2-2\beta\delta\beta ]F= 0,
\end{equation}
where we have neglected the term proportional to $(\delta\beta)^2$, since $\delta\beta$ is assumed to be a small perturbation to $\beta(\omega)$.  Now, if we multiply \eqref{Perturbation} by $F^*$, and integrate over $x$ and $y$, we obtain a formula for $\delta\beta$ in terms of $\delta\epsilon$
\begin{equation}\label{Beta}
\delta\beta= \frac{k_0^2}{2\beta_0}\int_{-\infty}^{\infty}\int_{-\infty}^{\infty}\delta\epsilon|F|^2 dxdy\Big/\int_{-\infty}^{\infty}\int_{-\infty}^{\infty}|F|^2 dxdy + i\frac{k_0^2}{2\beta}\mbox{Im}\tilde{\chi}^{(1)}_{xx},
\end{equation}
where $k_0\equiv k(\omega_0)$.  From \eqref{ModalAmplitude}, the modal amplitude $\tilde{A}(z,\omega-\omega_0)$ is found to satisfy
\begin{equation}\label{SimplerModalAmplitude}
i\frac{\partial\tilde{A}}{\partial z}+(\beta-\beta_0+\delta\beta)\tilde{A}=0,
\end{equation}
where we have approximated $(\beta+\delta\beta)^2-\beta_0^2$ by $2\beta_0(\beta-\beta_0+\delta\beta)$, since $\beta(\omega)\approx\beta_0$ when $|\omega-\omega_0|\ll\omega_0$.  Applying the same reasoning, we may expand $\beta(\omega)$ in a Taylor series around the carrier frequency $\omega=\omega_0$
\begin{equation}\label{TaylorSeries}
\beta(\omega) = \beta_0 + \beta_1(\omega-\omega_0) + \beta_2(\omega-\omega_0)^2/2,
\end{equation}
where the cubic and higher terms of $\omega-\omega_0$ are dropped in accordance with the assumption $|\omega-\omega_0|\ll\omega_0$.  Substitution of \eqref{TaylorSeries} into \eqref{SimplerModalAmplitude} immediately yields
\begin{equation}
i\frac{\partial\tilde{A}}{\partial z}+\left[\beta_1(\omega-\omega_0) + \frac{\beta_2}{2}(\omega-\omega_0)^2+\delta\beta\right]\tilde{A}=0.
\end{equation}
We now employ inverse Fourier transform to recover the modal amplitude in time domain
\begin{equation}
A(z,t) \equiv \frac{1}{2\pi}\int_{-\infty}^{\infty} \tilde{A}(z,\omega-\omega_0) e^{-i(\omega-\omega_0)t}d\omega.
\end{equation}
After employing inverse Fourier transform, $-i(\omega-\omega_0)\tilde{A}(z,\omega-\omega_0)$ is replaced by $\partial A(z,t)/\partial t$ and $-(\omega-\omega_0)^2\tilde{A}(z,\omega-\omega_0)$ is replaced by $\partial^2A(z,t)/\partial t^2$. Thus, the modal amplitude in time domain is found to obey
\begin{equation}\label{NonlinearSchrodinger}
i\frac{\partial A}{\partial z}+i\beta_1\frac{\partial A}{\partial t}-\frac{\beta_2}{2}\frac{\partial^2 A}{\partial t^2}+\frac{i\alpha}{2}A+\gamma|A|^2A=0,
\end{equation}
where $\alpha\equiv (k_0/n_e)\mbox{Im}\tilde{\chi}^{(1)}_{xx}(\omega_0)$ is the absorption coefficient that accounts for power losses of optical field inside the fiber, $n_e(\omega_0)\equiv \beta_0/k_0$ is the effective refractive index \cite{liu2009photonic} for the fundamental mode at frequency $\omega_0$, where $n_e$ takes values between the refractive indices of the core and cladding, $n_c\leq n_e \leq n_1$, and $\gamma$ is a nonlinearity parameter defined by
\begin{equation}\label{GammaFormula}
\gamma(\omega_0) \equiv \frac{3k_0}{8n_e}\chi^{(3)}_{xxxx}\int_{-\infty}^{\infty}\int_{-\infty}^{\infty}|F|^4 dxdy\Big/\int_{-\infty}^{\infty}\int_{-\infty}^{\infty}|F|^2 dxdy .
\end{equation}
Here $\beta_1\equiv \partial \beta/\partial \omega$ and $\beta_2\equiv \partial \beta_1/\partial \omega$ are related to the effective refractive index $n_e$ through
\begin{subequations}
\begin{align}
\beta_1&=\frac{1}{c}\frac{d(n_e\omega)}{d\omega}=\frac{n_e}{c}+\frac{\omega}{c}\frac{dn_e}{d\omega}\equiv \frac{1}{v_g},\\
\beta_2&=\frac{2}{c}\frac{dn_e}{d\omega}+\frac{\omega}{c}\frac{d^2n_e}{d\omega^2}.
\end{align}
\end{subequations}
Hence $\beta_1$ is the inverse of the effective group velocity $v_g$, and $\beta_2$ is a second-order dispersion parameter which accounts for the distortion of the wave envelope at different frequencies.  From \eqref{NonlinearSchrodinger}, we see that in the absence of dispersion effects, the transmitted power of the optical field at $z=L$ is given by $P=P_0e^{-\alpha L}$, where $P_0$ is the power of the optical field launched at $z=0$.  The absorption coefficient $\alpha$ for modern silica fibers is extremely small, typically around 0.2 dB/km at 1.55 $\mu$m.  Even a 10 dB/km loss corresponds to an absorption coefficient of only $\alpha \approx 2.5\times 10^{-5}$ cm$^{-1}$ \cite{agrawal2000nonlinear}. Hence, we may safely neglect the absorption coefficient for silica fibers.  Finally, we may introduce a frame of reference moving with the optical pulse at the group velocity $v_g$ by making the transformation $\tau \equiv t-z/v_g = t-\beta_1 z$, so that in the retarded frame, the wave equation for the modal amplitude becomes the nonlinear Schr\"{o}dinger equation
\begin{equation}
i\frac{\partial A}{\partial z}-\frac{\beta_2}{2}\frac{\partial^2 A}{\partial \tau^2}+\gamma|A|^2A=0,
\end{equation}
where $z$ plays the role of time and the retarded time $\tau$ plays the role of spatial variable in the conventional nonlinear Schr\"{o}dinger equation.

In deriving the nonlinear Schrödinger equation in an optical fiber, certain simplifications were made, including the assumption that ideal optical fibers with perfectly circular symmetry in the core cross-section maintain a stable polarization state along their length. However, in practical applications, this assumption may not always hold, as guided waves in optical fibers can be influenced by various environmental factors. Slight deviations from perfect circular symmetry due to nonuniform stress distribution or core-shape variations can introduce birefringence, allowing a single-mode fiber to support two nearly degenerate orthogonally polarized modes with slightly different propagation constants.

The degree of birefringence in such fibers is characterized by the difference between $\beta_x$ and $\beta_y$, the propagation constants of the HE$_{11}$ mode along the $x$ and $y$ axes, respectively. If the core cross-section remains unchanged throughout the fiber length, the fiber exhibits modal birefringence, quantified as $B_m\equiv |\beta_x-\beta_y|/k_0$ \cite{kaminow1981polarization}. Given a specific birefringence value, the phase difference between these orthogonally polarized modes reaches $2\pi$ after propagating a distance of $L_B\equiv 2\pi/|\beta_x-\beta_y|$ \cite{wolinski1999polarization}, referred to as the beat length. This beat length determines the periodicity of the polarization state along the fiber.

Apart from perturbative birefringence induced by environmental factors, some optical fibers are intentionally engineered to exhibit permanent birefringence. These so-called polarization-maintaining fibers \cite{ramaswamy1978polarization, payne1982development} rely on built-in birefringence mechanisms, such as elliptical core cross-sections or non-axially symmetric stress distributions surrounding the core, to ensure a stable polarization state. In fibers with elliptical core cross-sections, the modal birefringence strength is approximately $10^{-6}$, whereas fibers with an intentionally asymmetric stress distribution can achieve modal birefringence values as high as $B_m\approx 10^{-4}$ \cite{agrawal2000nonlinear}. Low-birefringence fibers ($B_m\approx 10^{-6}$) typically exhibit beat lengths of around $1\:$m, while high-birefringence fibers ($B_m\approx 10^{-4}$) have beat lengths shorter than $1\:$cm.

\section{General expression of homogeneous Hamiltonian}\label{B}
In this appendix, we present a reasoning of why only specific coupled-mode equations can be written in the form of nonlinear Schr\"{o}dinger equations. Let us consider a Hamiltonian $H(u_x,u_y)$ that is homogeneous in $u_x$ and $u_y$, which obey
\begin{equation}\label{homogeneous}
H(\lambda u_x,\lambda u_y)=H(u_x,u_y),
\end{equation}
where $\lambda$ is an arbitrary unit complex number. To simply the discussion, we assume that $u_x$ and $u_y$ are functions of only $z$. In order to fulfill the condition \eqref{homogeneous}, the Hamiltonian must be a combination of $|u_x|^2$, $|u_y|^2$, $u_xu_y^*$ and $u_yu_x^*$. Hence, up to quartic terms, we may write the Hamiltonian into the form $H=H^{(2)}+H^{(4)}$, where
\begin{subequations}
\begin{align}\label{Hamiltonian2}
H^{(2)}&=A|u_x|^2+B|u_y|^2+Cu_xu_y^*+Du_yu_x^*,\\
H^{(4)}&=E_1|u_x|^4+E_2|u_y|^4+F_1|u_x|^2u_xu_y^*+F_2|u_y|^2u_yu_x^*\nonumber\\
&+G_1|u_x|^2u_yu_x^*+G_2|u_y|^2u_xu_y^*+H|u_x|^2|u_y|^2\nonumber\\
&+I_1u_x^2u_y^{*2}+I_2u_y^2u_x^{*2}\label{Hamiltonian4}
\end{align}
\end{subequations}
As the Hamiltonian must be a real function of $u_x$ and $u_y$, $H^*(u_x,u_y)=H(u_x,u_y)$, we have further restrictions on the coefficients: both $A$, $B$, $E_1$, $E_2$, and $H$ are real numbers, where the remaining coefficients satisfy
\begin{equation}\label{restrictions}
D=C^*,G_1=F_1^*,G_2=F_2^*,I_2=I_1^*.
\end{equation}
Now, if we assume the coupled-mode equations can be written in the form of nonlinear Schr\"{o}dinger equations, Eqs.\:\eqref{Hamiltonian2} - \eqref{Hamiltonian4}, and Eq.\:\eqref{restrictions} immediately yield
\begin{subequations}
\begin{align}\label{GeneralNonlinearEquation1}
i\dot{u}_x&=\frac{\partial H}{\partial u_x^*}=Au_x+C^*u_y+2E_1|u_x|^2u_x+F_1u_x^2u_y^*+F_2|u_y|^2u_y\nonumber\\
&+2F_1^*|u_x|^2u_y+H|u_y|^2u_x+2I_1^*u_y^2u_x^*,\\
i\dot{u}_y&=\frac{\partial H}{\partial u_y^*}=Bu_y+Cu_x+2E_2|u_y|^2u_y+F_1|u_x|^2u_x+F_2u_y^2u_x^*\nonumber\\
&+2F_2^*|u_y|^2u_x+H|u_x|^2u_y+2I_1u_x^2u_y^*.\label{GeneralNonlinearEquation2}
\end{align}
\end{subequations}
As we can see in the primary article, the general form of coupled-mode equations when $u_x$ and $u_y$ are functions of only $z$ is
\begin{align}\label{GeneralCoupledModeEquations}
&i\frac{d u_j}{d z}+\frac{\xi_j}{2}u_j+a_j|u_j|^2u_j+b_j\left(2|u_k|^2u_j+u_k^2u_j^*\right)\nonumber\\
&+c_j\left(2|u_j|^2u_k+u_j^2u_k^*\right)+d_j|u_k|^2u_k=0,
\end{align}
Comparing Eqs.\:\eqref{GeneralNonlinearEquation1} - \eqref{GeneralNonlinearEquation2} and Eq.\:\eqref{GeneralCoupledModeEquations}, we have $\xi_x=-2A$, $\xi_y=-2B$,
\begin{equation}
  \begin{array}{l}
  \left\{
   \begin{array}{c}
   a_x=-2E_1,\\
   a_y=-2E_2,\\
   d_x=-F_2,\\
   d_y=-F_1,
   \end{array}
  \right.
  \left\{
   \begin{array}{c}
   b_x=-H/2=-2I_1^*,  \\
   b_y=-H/2=-2I_1,  \\
   c_x=-F_1^*=-F_1,  \\
   c_y=-F_2^*=-F_2.
   \end{array}
  \right.  \\ 
  \end{array}
\end{equation}
which implies that $F_1$, $F_2$, and $I_1$ are real numbers, and hence both $\xi_j$, $a_j$, $b_j$, $c_j$, and $d_j$ are real numbers, and obey
\begin{equation}\label{GeneralRestrictions}
b_x=b_y=b,d_y=c_x,d_x=c_y.
\end{equation}
As long as the condition \eqref{GeneralRestrictions} is fulfilled, the coupled-mode equations can be written in the form of nonlinear Schr\"{o}dinger equations, and the associated Hamiltonian has the form
\begin{align*}
&H=-\left[\frac{\xi_x}{2}|u_x|^2+\frac{\xi_y}{2}|u_y|^2+\frac{a_x}{2}|u_x|^4+\frac{a_y}{2}|u_y|^4+2b|u_x|^2|u_y|^2\right.\nonumber\\
&\left.+(c_x|u_x|^2+c_y|u_y|^2)(u_xu_y^*+u_yu_x^*)+\frac{b}{2}(u_x^2u_y^{*2}+u_y^2u_x^{*2})\right].
\end{align*}

\section{Generalized Lipkin–Meshkov–Glick Model}\label{C}

The Lipkin–Meshkov–Glick (LMG) model \cite{lipkin1965validity} is a cornerstone in the study of quantum many-body systems, originally introduced to model collective spin interactions in nuclear physics. Its generalized form extends the standard model by incorporating anisotropic interactions, external fields, and higher-order spin couplings, thereby providing a versatile framework to investigate cooperative phenomena, quantum phase transitions, and the interplay between quantum dynamics and classical rotational analogs. This appendix outlines the generalized LMG model, its Hamiltonian, and its connection to classical dynamics analogous to a rigid-body Euler top.

The standard LMG model describes a system of \(N\) spin-\(\frac{1}{2}\) particles interacting via infinite-range, uniform couplings. In the generalized version, the introduction of anisotropic coupling constants and external magnetic fields enriches the model’s dynamics. Its Hamiltonian is given by \cite{opatrny2018analogies}
\begin{equation}
H = -\frac{1}{N} \sum_{\alpha=x,y,z} \lambda_\alpha \left( S_\alpha \right)^2 - \sum_{\alpha=x,y,z} h_\alpha S_\alpha,
\label{eq:generalized_lmg}
\end{equation}
where $S_\alpha = \sum_{i=1}^N \frac{\sigma_\alpha^{(i)}}{2}$ with \(\sigma_\alpha^{(i)}\) being the Pauli matrices for the \(i\)-th spin, \(\lambda_\alpha\) denoting the coupling strengths along the three spatial directions, and \(h_\alpha\) the components of the external magnetic field. The factor \(1/N\) ensures a well-defined thermodynamic limit as \(N\to\infty\). The anisotropy in the parameters \(\lambda_\alpha\) allows exploration of regimes beyond the isotropic case where \(\lambda_x = \lambda_y = \lambda_z\). For example, setting \(\lambda_z = 0\) and \(\lambda_x=\lambda_y\) recovers a model with interactions confined to the \(xy\)-plane. In some generalizations, higher-order terms such as \((S_\alpha)^4\) are introduced, but here we focus on the quadratic form.

In the large-\(N\) limit, a semiclassical approach is often employed by mapping the collective spin operators to a classical spin vector \(\mathbf{S}=(S_x,S_y,S_z)\). Using the Poisson bracket relation \(\{S_\alpha,S_\beta\}=\epsilon_{\alpha\beta\gamma}S_\gamma\), the equations of motion become
\begin{equation}
\frac{dS_\alpha}{dt} = \epsilon_{\alpha\beta\gamma} S_\beta \left(\frac{2\lambda_\gamma}{N}S_\gamma + h_\gamma \right),
\label{eq:motion}
\end{equation}
where \(\epsilon_{\alpha\beta\gamma}\) is the Levi–Civita symbol. This form clearly mirrors the Euler equations for a rigid body, with the coupling constants \(\lambda_\alpha\) playing a role similar to the inverse moments of inertia.

The quadratic terms in Eq.~\eqref{eq:generalized_lmg} are responsible for generating spin squeezing—a quantum phenomenon in which the uncertainty in one spin component is reduced below the standard quantum limit, at the expense of increased uncertainty in the conjugate component. Spin squeezing is central to quantum metrology and provides insights into many-body entanglement. Two principal mechanisms are:

One-axis twisting (OAT) \cite{kitagawa1993squeezed}: When one quadratic term dominates (e.g., setting \(\lambda_z\neq0\) and \(\lambda_x=\lambda_y=0\) with \(h_\alpha=0\)), the Hamiltonian reduces to
\begin{equation}
H_{\rm OAT} \equiv -\frac{\lambda_z}{N} S_z^2.
\end{equation}
This generates a twisting dynamics via the time-evolution operator $U(t)\equiv\exp(i \lambda_z t S_z^2/N)$, which shears the uncertainty ellipse of an initial coherent state. The Kitagawa and Ueda’s squeezing parameter \cite{kitagawa1993squeezed}, $\xi^2\equiv (2/S)((\Delta S_{\boldsymbol{n}_{\perp}})^2)$, quantifies the reduction in quantum noise, with \(\xi^2<1\) indicating squeezing. For single-axis twisting, Kitagawa and Ueda \cite{kitagawa1993squeezed} showed that the optimal squeezing is achieved at a time scaling as $t \propto N^{-1/3}$. At this optimal time, the minimal squeezing parameter behaves as $\xi_{\min}^2 \propto N^{-2/3}$. Thus, as the number of particles increases, the amount of squeezing improves according to this \(N^{-2/3}\) law.

Two-axis counter twisting (TACT) \cite{kitagawa1993squeezed}: To achieve a more symmetric and efficient squeezing, one may employ a Hamiltonian that couples two spin components simultaneously. A typical form is
\begin{equation}
H_{\rm TACT} \equiv \chi\,(S_x^2 - S_y^2),
\end{equation}
or, up to a rotation, one that contains cross terms such as \(S_xS_y + S_yS_x\). The two-axis counter twisting yields an optimal squeezing that scales as \(\xi^2_{\rm TACT}\propto 1/N\), thereby outperforming single-axis twisting and approaching the Heisenberg limit for large \(N\). This indicates that the TACT scheme can—at least in principle—approach the Heisenberg limit, outperforming the OAT scheme.

In the semiclassical limit, the twisting driven by the quadratic terms corresponds to precession and nutation of the classical spin vector. In this picture, single-axis twisting deforms the spin’s phase-space trajectory uniaxially, whereas two-axis twisting produces biaxial distortions analogous to the complex rotational dynamics of an asymmetric Euler top.

Beyond squeezing phenomena, the generalized Lipkin--Meshkov--Glick (LMG) model offers a rich platform for investigating quantum phase transitions and critical phenomena. By tuning the parameters \( \lambda_\alpha \) and \( h_\alpha \), one can drive the system through distinct phases---for example, transitioning between ferromagnetic and paramagnetic states. In certain regimes, such as when one or more \( \lambda_\alpha \) vanish, the Hamiltonian becomes exactly solvable, thereby preserving its integrable character. In contrast, when the model is modified to incorporate additional degrees of freedom (for instance, by coupling multiple collective spins or introducing interactions that break the inherent all-to-all symmetry), it departs from integrability and can exhibit complex dynamics, including chaos \cite{zhang1988dynamical, kam2023coherent}. This evolution from an integrable regime to one displaying non-integrable behavior remains an active area of research, particularly in the context of classical--quantum correspondence.

\end{appendix}

\bibliography{references2}

\end{document}